\documentstyle[11pt,aasms]{article}
\lefthead{??}
\righthead{}
 
\newcommand{\kms}{\mbox{km s$^{-1}$}}

\def\lta{{\>\rlap{\raise2pt\hbox{$<$}}\lower3pt\hbox{$\sim$}\>}}
\def\gta{{\>\rlap{\raise2pt\hbox{$>$}}\lower3pt\hbox{$\sim$}\>}}

\lefthead{??}
\righthead{}
 
\begin{document}

\title {Spiral Galaxies with HST/NICMOS. I. Nuclear Morphologies,
Color Maps and Distinct Nuclei\footnote{Based
on observations with the NASA/ESA Hubble Space Telescope, obtained at
the Space Telescope Science Institute, which is operated by
Association of Universities for Research in Astronomy, Inc.\ (AURA),
under NASA contract NAS5-26555}}

\author{C. Marcella Carollo\footnote{Visiting Astronomer, The Johns
Hopkins University, Department of Physics and Astronomy, Baltimore MD
21210}} \affil{Columbia University, Department of Astronomy, 538
W.\ 120$^{th}$ St., New York, NY 10027}

\author{Massimo Stiavelli} \affil{Space Telescope Science Institute,
3700 San Martin Drive, Baltimore MD 21218}

\author{Marc Seigar\footnote{Joint
Astronomy Centre, 660 N. A'ohoku Place, Hilo, HI 96720} \affil{Sterrenkundig
Observatorium, Universiteit Gent, Krijgslaan 281, B-9000 Gent,
Belgium \\ Space Telescope
Science Institute, 3700 San Martin Drive, Baltimore MD 21218}}

\author{P. Tim de Zeeuw} \affil{Sterrewacht Leiden, Postbus 9513, 2300
RA Leiden, The Netherlands}

\author{Herwig Dejonghe} \affil{Sterrenkundig Observatorium,
Universiteit Gent, Krijgslaan 281, B-9000 Gent, Belgium}

\begin{abstract}
This is the first of two papers where we present the analysis of an
HST {\tt NICMOS-Cam2} near-infared (NIR) snapshot survey in the F160W
($H$) filter for a sample of 78 spiral galaxies selected from the UGC
and ESOLV catalogs. For 69 of these objects we provide nuclear color
information derived by combining the $H$ data either with additional
{\tt NICMOS} F110W ($J$) images or with $V$ {\tt WFPC2}/HST data.  \\
Here we present the NIR images and the optical-NIR color maps. We
focus our attention on the properties of the photometrically-distinct
`nuclei' which are found embedded in most of the galaxies, and provide
measurements of their half-light radii and magnitudes in the $H$ (and
when available, in the $J$) band. We find that: {\it (i)} In the NIR,
the nuclei embedded in the bright early- to intermediate-type galaxies
span a much larger range in brightness than the nuclei which are
typically found embedded in bulgeless late-type disks: the nuclei
embedded in the early- to intermediate-type galaxies reach on the
bright end values up to $H_{AB} \sim-17.7$ mags; {\it (ii)} Nuclei are
found in both non-barred and barred hosts, in large-scale ($\gta$ 1
kpc) as well as in nuclear (up to a few 100pc) bars; {\it (iii)} There
is a significant increase in half-light radius with increasing
luminosity of the nucleus in the early/intermediate types (a decade in
radius for $\approx 8$ magnitudes brightening), a correlation which
was found in the $V$ band and which is also seen in the NIR data; {\it
(iv)} The nuclei of early/intermediate-type spirals cover a large
range of optical-NIR colors, from $V-H\approx$ -0.5 to 3.  Some nuclei
are bluer and others redder than the surrounding galaxy, indicating
the presence of activity or reddening by dust in many of these
systems; {\it (v)} Some early/intermediate nuclei are elongated and/or
slightly offset from the isophotal center of the host galaxy. On
average, however, these nuclei appear as centered, star-cluster-like
structures similar to those which are found in the late-type disks.
\end{abstract}

{\it subject headings}: galaxies: formation - galaxies: evolution -
galaxies: structure - galaxies: nuclei - galaxies: spirals - galaxies: bulges

\section{Introduction}

\noindent
The nuclear regions of spiral galaxies may hold important information
concerning the growth of bulge-like structures within the disks, and
more generally the secular evolution of the disks themselves.  Little
was known about the $\sim 10$-100 pc scale regions of spiral galaxies
before we started our HST program, a cycle 6 {\tt WFPC2} F606W ($V$)
survey of 75 mostly early- and intermediate-type disk galaxies
selected randomly from a complete sample of 134 galaxies (Carollo et
al.\ 1997; Carollo, Stiavelli \& Mack 1998; Carollo \& Stiavelli 1998;
Carollo 1999). The random selection is due to the snapshot nature of
the program. The unprecedented angular resolution of the HST optical
images allowed probing the bulges down to absolute $V$ magnitudes $M_V
\sim -14$, with half-light radii $R_e$ as small as $\sim 100$pc, and
unveiled an astonishing richness of structure in the nuclear regions
of $\approx 60\%$ of the sample: bars, spiral-like dust lanes,
star-forming rings or spiral arms. Even more surprisingly, a similar
fraction of objects contains photometrically-distinct nuclear compact
sources (hereafter `nuclei'). While a few of these nuclei appear
point-like in the {\tt WFPC2} images, many are resolved with $R_e
\approx 0.1''-0.2''$, corresponding to linear scales of up to $\approx
20$ pc.  The nuclei were identified primarily by visual inspection of
the images, since they appear as regular, smooth structures
significantly brighter than their surroundings. For the
subsample of objects with a light distribution smooth enough to allow
the derivation of an isophotal fit, the morphological identification
of the nuclei was confirmed by inspecting the surface brightness
profiles. The nuclei embedded in those hosts whose dusty/star-forming
morphology did not allow us to obtain a meaningful isophotal fit are
typically much brighter than the nuclei embedded in the smoother
hosts, so that their identification as photometrically-distinct
systems remains unambiguous.  Carollo et al.\ (1997,1998) presented a
collection of $V$-band images and light profiles which clearly showed
the photometric decoupling between the nuclei and their underlying
host galaxies.  Faint nuclei embedded in bright hosts with steep
nuclear stellar cusps would remain undetected (this is discussed
quantitatively in the references above).

The relation between the nuclei and their galactic surroundings is
unclear. In principle, the nuclei could be triggers for ---and/or
left-overs of--- the mechanisms which produce central structure within
the disks, and could thus provide additional constraints for
identifying such mechanisms.  Interestingly, the {\tt WFPC2} survey
has shown that many of the nuclei at the faint-end of their luminosity
distribution ($-9 \gta M_V \gta -12$) are often hosted by
pseudo-bulges with an exponential ---rather than $R^{1/4}$-law---
fall-off of the light distribution (hereafter exponential bulges;
Courteau, de Jong \& Broeils 1996 and references therein; see also the
review by Wyse, Gilmore \& Franx 1997).  In fact, every exponential
bulge in our {\tt WFPC2} sample hosts a distinct nucleus in its center
or slightly offset from it.  Distinct central structures significantly
different from the massive, $R^{1/4}$-law elliptical-like spheroids
are rather common in the intermediate-type Sb-Sc disks.  In the
latter, in addition to the exponential bulges, pseudo-bulges have been
found with cold kinematics (Kormendy 1993) or a peanut-shape
morphology (Kuijken \& Merrifield 1995; Bureau \& Freeman 1999).  The
presence of the nuclei in the centers of the exponential bulges does
not necessarily imply a causal connection between nuclei and
bulges. At least at the heuristic level, however, such a connection
may be indirectly supported on a theoretical basis, since evolutionary
scenarios for bulge formation advocate for example the dissolution of
a progenitor bar by means of the accumulation of a $\sim 1\%$ central
mass (Pfenniger \& Norman 1990; Hasan, Pfenniger \& Norman 1993;
Pfenniger 1993; Norman, Sellwood \& Hasan 1996, and references
therein).

A first step toward understanding the nucleus-host relationship is to
establish the properties of the nuclei as a function of the Hubble
type and other physical properties of their parent galaxies.  The
findings based on the optical {\tt WFPC2} data could potentially be
severely affected by large amounts of dust and of recent star
formation that may be present in the central regions of spirals. To
secure results less influenced by these sources of `pollution', we
have followed up the {\tt WFPC2} survey with a {\tt NICMOS} Camera-2
F160W ($H$; in some cases, also F110W, i.e., ``broad'' $J$)
snap-shot survey for the same sample of galaxies.

The analysis of our {\tt NICMOS} survey is reported in this and in a
companion paper (Seigar et al.\ 2001, paper II).  In this paper we:
{\it (i)} Introduce the {\tt NICMOS} sample and describe our technique
of data reduction; {\it (ii)} Describe the NIR images and the
optical-NIR color maps of spirals at HST resolution; {\it (iii)}
Detail the methodology used to measure the sizes and luminosities of
the nuclei from the {\tt NICMOS} images, and describe the tests
performed to assess the systematic errors on these measurements; {\it
(iv)} Discuss the statistical NIR and optical-NIR properties of the
nuclear regions of spirals, including the distinct nuclei, as a
function of Hubble type.  In paper II we present the isophotal fits to
the galaxy nuclear surface brightness distribution, and the analytical
fits to these profiles. Paper II also reports the derived nuclear
stellar cusp slopes and the optical/NIR color profiles (for those
galaxies for which both {\tt WFPC2} and {\tt NICMOS} images are
available).  The data and measurements presented here and in paper II
are used in an additional paper (Carollo et al.\ 2001) to discuss in
more detail the nucleus-bulge connection. There we present the
optical/NIR color-distribution, color-magnitude and color-color
diagrams for the $R^{1/4}$ and exponential bulges and for their
nuclei, and discuss the possible implications for the formation and
evolution of the central regions of early/intermediate-type disk
galaxies.

In \S 2 we describe the {\tt NICMOS} sample, the observations, the
data reduction procedure and the approach adopted to derive the radii
and luminosities of the nuclei.  In the same section we present the
tests that we performed to assess the accuracy of the measurements for
the nuclei, and show the optical-NIR or NIR-NIR color maps for the
target galaxies. We discuss in \S 3 the sizes and magnitudes of the
nuclei in the NIR, their colors, as well as the structure and color of
their surrounding environments.  We sumarize in Section 4.  Consistent with
our earlier work, we adopt $H_0=65$ km/s/Mpc throughout this paper.

\section{Sample and Data Analysis}

\noindent
The {\tt NICMOS} sample presented here constitutes a random selection
from the sample of 134 galaxies discussed in Carollo et al.\ (1997,
1998).  The objects were extracted from the UGC catalog for the
northern hemisphere (Nilson 1973) and the ESOLV catalog for the
southern hemisphere (Lauberts \& Valentijn 1989) for having angular
diameter larger than 1 arcmin, redshift $<$ 2500~\kms, inclination
angle smaller than 75 degrees, and (non-barred) morphological type
from Sa to Sbc.  A few objects that did not satisfy these selection
criteria (e.g., NGC 1365 and NGC 4565) were however included in the
final sample in substitution of objects which we were not allowed to
observe due to the HST rules on duplication of targets.  A posteriori,
the {\tt WFPC2} images showed that the RC3 catalog (de Vaucouleurs et
al.\ 1991) gives generally a better description of the galaxy
types. Therefore, consistent with our previous {\tt WFPC2} work, we
adopt in this analysis the RC3 Hubble types (with the numerical
conversion given in Table 2 of RC3:  0=S0a, 1=Sa, 2=Sab,
$\ldots$, 9=Sm).

\subsection{The Dataset and the Basic Data Reduction}

\noindent
All {\tt NICMOS} observations were performed with fine-lock guiding.
78 of the 134 systems were observed in snap-shot mode with {\tt
NICMOS} Camera-2 in F160W during the period June 1997 to February
1998.

Nine objects,  NGC1365, NGC2903, NGC3031, NGC4536, NGC4565,
NGC4569, NGC6217, NGC7479 and NGC7742 were observed with {\tt WFPC2}
in the context of other programs; consequently, we were not allowed to
`duplicate' the optical observations for these 9 systems. The archival
images for these galaxies, however, were not taken in the F606W filter
which we had chosen for our own {\tt WFPC2} survey. Given the
arbitrary assumptions that would underlie the conversion from the
``archival''-F160W into the F606W-F160W color (a conversion necessary
for studying these systems together with the other objects of our
sample), we have not included these 9 galaxies in the statistical
study presented in this paper, to guarantee the homogeneity of the
results. The $H$ images of these systems at HST resolution are
presented in Appendix A.  The remaining 69 objects are considered in
this study.  At the time of the {\tt NICMOS} Phase II submission, no
{\tt WFPC2} data were available either in the archive or as part of
our snapshot program for 46 of the targets. For these objects a F110W
image was also included in our {\tt NICMOS} observing plan; 41 of
these 46 systems were actually observed in our {\tt NICMOS} survey
with both the F160W and the F110W filter. These F110W and F160W
exposures consist of two MULTIACCUM sequences with STEP256 and
NSAMP=10, for a total exposure time of 256 seconds per filter.  For 30
of these 46 galaxies {\tt WFPC2} F606W images were obtained, as part
of our WFPC2 snapshot program, after the {\tt NICMOS} phase II
deadline. Thus, images in the three filters F606W, F110W and F160W,
are available for these 30 systems.  For the remaining targets F606W
{\tt WFPC2} images were already in hand at the time of the {\tt
NICMOS} phase II deadline. For these objects we acquired three F160W
MULTIACCUM sequences with STEP256 and NSAMP=10, which results in a
total exposure time of 384 seconds.  Images in both F160W and F606W
are available for a total of 58 galaxies. Table 1 lists the 69
galaxies considered in this study together with the list of filters in
which they have been observed, the galaxy distance in Mpc,
morphological type, apparent total $B$ magnitude, $B$ extinction, as
well as the information on whether an $R^{1/4}$ or an exponential
bulge is present, and the type of central spectrum observed at
ground-based resolution.

for
these objects, including whether they were also observed within our
program in F110W and/or in F606W.

The {\tt NICMOS} images were reduced with the standard pipeline
software, using the on-orbit flats and darks identified in the HST
archive as the best-reference frames. The various {\tt NICMOS}
image-anomalies were corrected on a case-by-case basis. All images
were affected by the so-called `pedestal' anomaly, a time-variable
bias of unclear origin. In most of the images the pedestal was removed
using the {\tt NICMOS} Pedestal Estimation Software (developed by
R.P. van der Marel).  In about 10$\%$ of the images the pedestal
effects were very severe, and they had to be removed iteratively.  In
some galaxies (ESO240G12, ESO404G3, ESO443G80, ESO548G29, ESO549G18,
ESO572G22, IC1555, NGC1345, NGC1800, NGC3455, NGC4806, NGC4980,
NGC7188 and NGC7241), even after removal of the pedestal, a constant
gradient across the chip remained visible. This was removed by
subtracting a linear fit to the edges of the chip, using the
`imsurfit' routine in {\tt IRAF}.  In about 30$\%$ of the images
additional anomalies such as shading and bias jumps/bands were
present. These were also removed from the images before performing the
measurements.

The combined images were flux-calibrated adopting the pipeline flux
calibration zeropoints,  ZP$_{F110W}$ = 23.248 for the F110W
filter and ZP$_{F160W}$ = 23.110 for the F160W filter (AB system, Oke
\& Gunn 1983).  Magnitudes for our WFPC2 data were originally given in
``Vegamag''; we converted them to the AB system by using V$_{\rm
606,AB}$ = V$_{\rm 606,Vega}$+0.117.  The conversion between AB and
Vegamags was computed by us using the Synphot synthetic photometry
package (in IRAF) with the spectrum of Vega as input.

$V-H$ color maps were derived for the {\tt NICMOS} galaxies for which
{\tt WFPC2} F606W images were also available, including 8 objects for
which the optical images were obtained after the publication of the
{\tt WFPC2} results (asterisks close to the galaxy name in Table 1).
To account for the difference in Point Spread Function (PSF) between
the $H$ and $V$ images, the images in a given band were convolved with
the PSF of the other band.  All PSFs were obtained with TINYTIM (Krist
1997).  The adopted extent of the PSF images generated by TinyTim was
3$''$. The PSF FWHM was 0.127$''$ for the F160W filter, and 0.091$''$
for the F110W filter, respectively; a $5\times 5$ sub-pixel scale was
used.  The $V$ and $H$ images were aligned after sub-pixel rebinning
before computing the color maps. For the 11 galaxies for which the
only additional images available were those in the F110W filter, we
computed the $J-H$ color maps following the same approach just
described for the optical-NIR maps.

The identification of the nuclei in the NIR images was based primarily
on visual inspection of the images, and was checked on the radial
light profiles for those systems for which an isophotal fit could be
derived (paper II). Similar to what was done for the visual {\tt
WFPC2} data, the nuclei were selected based on the criterion that
their compact, smooth, regular-morphology excess of light was
well-isolated from the remaining circum-nuclear structure.  In Figure
1 a we show the $H$ images and $V-H$ color maps for 38 of the 39
galaxies that host a distinct nucleus and that were imaged both in
F160W and F606W (the image of the 39$th$ object, ESO549G18, had a S/N
sufficient for measuring the parameters of the nucleus, but not for
studying the underlying galaxy); in Figure 1 b we show instead the $H$
and $J-H$ images for the 5 galaxies which host a distinct nucleus for
which the only other filter available in addition to F160W is F110W.
Figures 2 a and b show the similar images for those galaxies which
either do not host a distinct nucleus, or in which the presence of a
nucleus is uncertain. In Figures 3 a and b we show the $H$-band light
profiles of two galaxies representative of the two subsets presented
in Figures 1 and 2, respectively: the first galaxy hosts a distinct
nucleus embedded in an exponential pseudo-bulge (panel a; NGC 2082);
the other galaxy hosts an $R^{1/4}$-law bulge (panel b; NGC 2460). In
the NGC 2082 the photometrically-distinct nucleus is clearly
identifiable in the $H$ light profile inside the innermost few tenths
of arcsecond, similarly to what was found in the visual band. In
contrast, in the NIR as in the $V$ band, the $R^{1/4}$-law bulge shows
the steep luminosity profile typical of this class of objects (Carollo
et al.\ 1998; Carollo \& Stiavelli 1998; paper II).

\subsection{Radii and Luminosities of the Nuclei in the NIR}

\noindent
The NIR measurements of the magnitudes and radii of the nuclei are
affected by the uncertainty on the light contribution from the
underlying galaxy.  To quantify the luminosities of the nuclei in the
NIR we adopted the three approaches previously optimized to extract
similar information from the {\tt WFPC2} visual images, namely: (1) we
fitted the contribution from the nucleus with a Gaussian, assuming
that the underlying galactic light continuum is well represented by
the asymptotic value of the Gaussian wings; (2) we modeled the
underlying galactic light with a fourth-order polynomial using the
task ``fit/flat" in {\tt ESO/MIDAS}, subtracted this model from the
image, and attributed the excess counts to the nucleus; (3) we
subtracted from the original frame an iteratively smoothed and
nucleus-masked version of it, and performed the measurement on the
resulting image containing only the nucleus.  These three methods are
complementary: they sample the galactic background in different
regions, and hence their combined use allows us to estimate the
uncertainty in the derived values contributed by the poorly
constrained underlying galactic light.  As done for the optical
measurements, we adopted as final the averages of, and as errorbars
the mean of the differences between, these three estimates for the
luminosity of the central sources.  The sizes of the compact sources
were taken equal to the FWHM of the best fitting Gaussians of method
(1), corrected for the instrumental width of {\tt NICMOS} assuming
that the intrinsic profile of the nucleus is described by a Plummer
law (see Eq. 2).  Consistent with the {\tt WFPC2} study, in the
following we adopt for the half-light radii of the nuclei the ratio
between half-light radius and FWHM valid for a Plummer law (see
below). This gives a better representation than, e.g., a Gaussian
shape, given that all compact sources show wings in excess of a Gaussian. An
assessment of the systematic uncertainties associated with the adopted
radii and luminosities of the nuclei is given in Appendix B.

In our {\tt WFPC2} study of the nuclei (Carollo et al.\ 1997; Carollo
et al.\ 1998), for those galaxies with an isophotal/analytical fit
available, we also estimated the magnitudes and luminosities of the
nuclei by using two additional approaches: {\it (a)} We adopted the
best fitting (exponential) models of the bulges to describe the
underlying galaxian light, subtracted these models from the images,
and attributed the residual nuclear counts to the nuclei; {\it (b)} We
fitted the entire radial light profiles, including the innermost
points, with multi-component analytical laws representing the bulge
with an exponential law and the nucleus with either an $R^{1/4}$-law,
or a Plummer law, or a modified Hubble profile.  These alternative
approaches provided results in agreement with ---and well within the
range spanned by--- those obtained using the methods (1), (2), and (3)
discussed above. To guarantee a homogeneous and self-consistent set of
measurements, however, in our {\tt WFPC2} analysis we only used the
results obtained from the latter three methods, since the additional
two approaches could only be applied to a subset of the sample. Due to
the limited radial extension of the {\tt NICMOS} data, no bulge/disk
decomposition was performed on the NIR data, and therefore no
exponential fits are available for the bulges in $H$ or $J$. Thus,
these additional tests could not be performed on the NIR
profiles. Nonetheless, the results of the tests conducted on the {\tt
WFPC2} data give us confidence that also for the {\tt NICMOS}
measurements the error bars derived from the ``primary'' methods (1),
(2) and (3) do reliably represent the uncertainty associated with
disentangling the unknown contribution from the underlying galaxy.

Finally, we computed the location of the nuclei by measuring their
centroids, and the centers of their host galaxies by measuring the
centers of the isophotes computed at $\approx 1''$ (the precise radius
was slightly increased/decreased in some galaxies to avoid the effects
of prominent structures).  The elongation of the nuclei was
established by visual inspection.

\section{Results and Discussion}

\subsection{The Central Regions of Spirals in the NIR}

\noindent
A large structural complexity is observed in the $H$ band in the
central regions of many spiral galaxies; on the other hand, it is
clear that the {\tt NICMOS} images provide a much cleaner picture of
the underlying galaxy structure as compared to the heavily
dust-obscured {\tt WFPC2} visual images. In the extreme cases of NGC
3067, NGC 4750 and NGC 6239, the {\tt NICMOS} images unveil the
location of the true galaxy center, which we had been unable to
identify in the {\tt WFPC2} images due to strong dust features and
star formation. The {\tt NICMOS} images have revealed in two of these
galaxies (NGC 4750 and NGC 6239) the presence of a distinct compact
nucleus, which subsequently, with the acquired knowledge of its
precise location, we have been able to identify and parametrize also
in the {\tt WFPC2} images. The optical measurements for these nuclei
were not reported in our previous papers; these measurements are
reported here in Table 2, identified by the ``$\dagger$'' symbol. The
possibility of better ``penetrating'' the dust in the NIR with respect
to the optical data has also led to the identification, based on the
visual inspection of the $H$ images, of nuclear bar-like features in
ESO499G37, NGC 4384, NGC 4750, NGC 5448, NGC 5806, NGC 7280
(identified by `S' ---standing for small--- in the column `Bar' in
Table 3), and possibly also in IC4390, IC 5271, IC 5273, NGC 289, NGC
2339, NGC 3455, NGC 3898, NGC 3949, NGC 4102, NGC 4219, NGC 4806, NGC
5678, NGC 6810, NGC 7177, NGC 7188 and NGC 7690 (these ambiguous cases
are identified by 'S?'  in Table 3).  In IC 4390, IC 5271, NGC 3455,
NGC 3949, NGC 4219, and NGC 5448 the nuclear bar-like features become
evident in the $V-H$ or $J-H$ maps by their distinct colors (either
redder or bluer than the surroundings).  Consistent with the studies
which have looked for nuclear bars in the context of the feeding of an
active non-thermal nucleus, we find that even in the less
dust-sensitive NIR images the nuclear bar-like features are not
ubiquitous.  Interestingly, spiral structure down to the
center/nucleus is observable in the $H$-band images of several
systems.  In some galaxies,  ESO205G7, NGC3928, NGC5678, a
small-scale ring is observed around the centre/nucleus; these rings
are likely not an artefact of PSF mismatch (since changing the sizes
of the PSFs by 10\% does not affect the result), and are likely due to
circum-nuclear dust or star formation.  

In summary, the NIR images, coupled with the optical images, show that
there is no galaxy in our sample that is not ``polluted'' either by
dust or by recent star formation or both in its nuclear/circum-nuclear
region: all galaxies contain some nuclear/circum-nuclear feature
overlying the smooth/old stellar populations. Excluding a few objects
for which the signal-to-noise ratio of the NIR images is too low to
accurately measure the underlying light outside the bright nucleus,
the galaxies in our sample can be broadly grouped into a four classes,
based on their nuclear NIR- and optical/NIR-color morphology:

\smallskip

\noindent
{\sl Class 1. Galaxies with Concentrated Nuclear Star Formation Mixed with
Dust.}  Judging from the $H$ images and the optical-NIR color maps,
many galaxies (ESO 498 G5, ESO 499 G37, IC 4390, IC 5273, NGC 406, NGC
986, NGC 1483, NGC 1688, NGC 1800, NGC 2082, NGC 2566, NGC 2964, NGC
3067, NGC 3928, NGC 4102, NGC 4219, NGC 4384, NGC 5678, NGC 5806, NGC
6000, NGC 6329, NGC 6951, NGC 7217, NGC 7259) host in their innermost
few-hundred pc$^2$ concentrated star formation in the form of compact
blue knots, intermixed with strong red dust features. With the
exception of NGC 2566, NGC 3928, NGC 7217, and NGC 7259, where the
star forming and dust features have a regular spiral morphology, and
of NGC 6951, where they form a circum-nuclear ring, in all the
remaining objects the nuclear star formation and dust have a rather
irregular morphology. Only some of these systems have been found to
host a nuclear bar or a central nucleus, so that the presence of these
features is not a unique characteristic of this family of objects. All
but four (ESO 499 G37, IC 5273, NGC 1688, NGC 1800) of these galaxies
are early-to-intermediate type spirals in the Hubble range between Sab
and Sbc.  From these data it seems that star formation in the central
regions of early-type spiral galaxies does not occur in the rare,
transient, and intense ``starburst mode'': low-level star formation
appears to be common in these systems. A question to answer is whether
this low-level star formation represents the low-luminosity end of the
starburst phenomenon, or rather the fading evolutionary descendant of
a major starburst itself (Heckman 1997).

\noindent
{\sl Class 2. Galaxies with Diffuse Blue Nuclear Regions.}  Several systems,
namely ESO 240 G12, NGC 972, NGC 1345, NGC 1398, NGC 2339, NGC 2748,
NGC 3949, NGC 4527, NGC 5188, NGC 5377, NGC 5448, NGC 6810, NGC 7177,
and NGC 7690 also host significant amounts of circum-nuclear dust
mixed with rather blue surroundings; the latter however have a more
diffuse appearance than the knot-like features typical of the previous
group. For example, in NGC 5377 the blue region is embedded in a
completely-obscuring dust torus and has the morphology of regular disk
surrounding the distinct central nucleus. Among these systems there
are a few (NGC 5377, NGC 5448, NGC 6810) which are known to host an
AGN; for them, it is possible that the blue regions are related to
non-thermal activity from the central engine.

\noindent
{\sl Class 3. Galaxies with Regular Nuclear/Circum-Nuclear Dust.} There are 
some galaxies that do not show evidence for star formation (with the
possible exclusion of the nucleus), but host in their central regions
strong dust lanes with a regular, well-defined spiral or ring
morphology. These systems are ESO 205 G7, NGC 488, NGC 772, NGC 2460,
NGC 3259, NGC 3277, NGC 4750, NGC 6384, NGC 7013.  Some of these
objects show some ionized-gas emission on ground-based scales (Table
1), but this must be distributed smoothly in the central region not to
show up clearly in the images in the form of identifiable
structures. Apart from being all early-to-intermediate type systems
(S0a to Sbc), these galaxies show no other characterizing feature:
some belong to the group of Figure 1 and thus host a distinct nucleus,
others belong to the group of Figure 2; some have an exponential
bulge, others host instead an $R^{1/4}$ bulge. The nuclei which are
embedded in some of the galaxies in this class tend to have relatively
red $V-H$ colors, ranging from 1.3 to 2.1 magnitudes.

\noindent
{\sl Class 4. Galaxies with Irregular Nuclear/Circum-Nuclear Dust.}  Several
more objects in the sample also show a quiescent, non-starforming
morphology with strong dust features, but the latter are distributed in a
much more irregular and patchy way (although in some cases, in addition to
the irregular patches, spiral-like dust is also present). These
objects are IC 5271, NGC 289, NGC 1892, NGC 2196, NGC 2344, NGC 2758,
NGC 3455, NGC 3898, NGC 3900, NGC 4806, NGC 4980, NGC 5985, NGC 6340,
NGC 7162, NGC 7188, NGC 7280, NGC 7421, NGC 7513. Similar to the
case above, some of these systems host a central
distinct nucleus, others do not; with the exception of NGC 1892, they are
early-to-intermediate type systems, with Hubble types between S0 and
Sc. The $V-H$ colors of the four nuclei detected in this class of
systems range from 1.4 to 3.1 magnitudes; the latter high $V-H$
value clearly indicates the presence of a significant amount of dust
even on the nucleus itself. In several of these systems the
presence of a nucleus is uncertain, due to the dust which obscures the
central region.

\smallskip

Additional information is concisely reported in Table 3 where we
summarize, object by object, the observed properties of the
circumnuclear $V-H$ or $J-H$ light distribution.

\subsection{The Nuclei}

\noindent
Table 2 lists the F160W sizes and luminosities for the nuclei detected
in the {\tt NICMOS} sample, measured as described in \S 2.2. When
available, the sizes and luminosities of the nuclei in the F110W
passband are also listed, together with, for an easy comparison, the
F606W measurements published in Carollo et al.\ (1997, 1998) for the
{\tt NICMOS} galaxies observed within our program also with {\tt
WFPC2}. The F606W properties of the nuclei for the 8 {\tt NICMOS}
targets which were not included in our {\tt WFPC2} papers due to the
late acquisition of the {\tt WFPC2} images (\S 2.2) are also given in the
Table. The comparison of the sizes derived from the {\tt WFPC2} and
F110W images with those derived from the F160W images is shown in
Figure 4.  The radii measured with {\tt NICMOS} and {\tt WFCP2} are in
good agreement within the error bars.

Compact star-cluster-type nuclei in bulgeless, weakly- or non-active
late-type disks have been the subject of a few detailed studies
(Kormendy \& McClure 1993; Phillips et al.\ 1996; Matthews \&
Gallagher 1997; Lauer et al.\ 1998). In the visual band, these
late-type nuclei have been found to cover a very narrow range in
luminosity ($M_{I} \sim -11 \pm1$) despite the large range in
luminosity of their host galaxies (from $M_{V}=-15.1$ to
$M_{V}=-18.2$; Matthews et al.\ 1999).  Our {\tt WFPC2} and {\tt
NICMOS} surveys show that photometrically-distinct nuclei are very
frequent also in intermediate-type spirals. In contrast with the four
nuclei in our sample embedded in the late-type galaxies, which lie
typically `naked' in the centers of the disks (as reported also by
Matthews et al.\ 1999), in the earlier-types the nucleus is embedded
either in a smooth pseudo-bulge with an exponential radial light
profile, or, in the case of the brightest nuclei ($-12\gta M_V \gta
-16$), in a complex structure of strong dust lanes and star-forming
knots in irregular, spiral or ring-like patters typically still
observable also in the NIR (see also Carollo et al.\ 1997; Carollo et
al.\ 1998; Devereux et al.\ 1997).  Furthermore, the new data show
that:

{\it (i)} Within our sample, in all cases for which a nucleus was
detected in the {\tt WFPC2} images and for which the {\tt NICMOS} data
are available, the nucleus is also clearly seen in the NIR: this
suggests that all nuclei, independent of Hubble type, are relatively
massive, given that the NIR light around 1.6$\mu$ is a rather good
tracer of the stellar mass.  In the NIR as much as in the visual
passband, however, the nuclei embedded in the bright early- to
intermediate-type hosts span a much larger magnitude range than the
nuclei embedded in the fainter late-type disks, as shown in Figure 5
a, where the absolute $H_{AB}$ magnitude of the nuclei is plotted
against their half-light radius derived in the same passband: The four
nuclei embedded in the Scd and later-type disks have nuclear
properties quite similar to those reported in the quoted literature
for other late-type nuclei, and are therefore likely representative of
that category of objects.

{\it (ii)} In our sample we find nuclei that are either embedded
(e.g., NGC 6000) or not embedded (e.g., NGC 2964) in large-scale
($\gta$ 1 kpc) bars. We also find that some of the nuclei sit within
smaller bars, with scales of up to a few 100 pc (e.g., ESO499G37).
These small-scale bars could be particularly important within the
context of bar-driven secular evolution scenarios for bulge formation,
since they might provide a mechanism for bringing the disk material
from the outer regions down to the pc scales in the shallow
gravitational potential of the progenitor disks (Shlosman 1994;
Shlosman \& Robinson 1995; Shlosman, Begelman \& Frank 1990; Friedli
\& Martinet 1993). The nucleus/nuclear-bar connection is however not a
straightforward one since, as stressed above, nuclear bars are not a
universal feature of nucleated spiral galaxies.

{\it (iii)} The brightest nuclei embedded in the
early/intermediate-type galaxies tend to have larger radii than the
fainter nuclei. An increase of radius with increasing luminosity of
the compact nucleus in the early/intermediate types is detected in the
$H$ band (linear correlation coefficient of -0.76, corresponding to a
probability $> 99.9\%$). This was already noticed in the F606W
passband (see Figure 5 b, where the absolute magnitude versus
half-light radius relation for the nuclei is given for the $V$ band,
for a comparison with Figure 5 a, where the same relation is given the
$H$ band).  The same finding in the NIR, however, strengthens the
reliability of this result, given the much simpler circumnuclear
background structure in the $H$ band as compared to the complex one in
which (especially the brightest) nuclei were found embedded in the
visual images.  The measurements of the half-light radii of the nuclei
have a rather large uncertainty, mostly because at the distance of our
sources the extended light component of the nucleus must be separated
from the wings of the PSF (see also Matthews et al.\ 1999). The
galaxies of our sample are however restricted to $cz<2500$ km/s, which
ensures a rather high degree of homogeneity in the derived set of
measurements.  In contrast with the simulations, which showed a modest
brightening of the nucleus corresponding to a modest increase in
radius, the nuclei observed in the centers of early/intermediate-type
spirals span an entire decade in radius, over which they show an
increase in luminosity of about eight magnitudes. It is therefore
likely that a substantial part of this correlation is driven by a real
physical effect.

{\it (iv)} The resolution of the data is not appropriate to detect
color variations within individual nuclei, but it is adequate to
compare the nuclei to their surroundings.  The early/intermediate-type
spirals' nuclei are found to cover a large range of colors, from
$V-H\approx$ -0.5 to 3.  Previous HST studies on the colors of the
central regions of spirals report very red (dusty) centers for the
early-type bulges (Peletier at al.\ 1999), and rather blue colors for
the distinct nuclei of late-type disks (Matthews et al.\ 1999).  The
few late-type nuclei in our sample show a slight tendency for having
bluer colors than the nuclei of the earlier-type hosts, as is shown in
Figure 6, where the $V-H$ color of the nuclei is plotted against the
RC3 Hubble type of the host galaxy. Figure 6 also shows however that
the spread in $V-H$ color is rather large within the
early/intermediate types themselves: The early/intermediate nuclei
have an average $V-H \approx 1.44$ with a sigma of about 0.82
magnitudes.  In the Figure, the nuclei are identified with different
symbols according to the galaxy-class to which their hosts belong
(among the four described above; symbols are explained in the figure
caption). No significant trend is observed with galaxy-class, with
only a possible hint for the nuclei embedded in galaxies with regular
nuclear/circum-nuclear dust to be slightly redder than the average.  A
comparison of each nucleus with its own galactic surroundings shows
that the nuclei are in some cases redder and in others bluer than
their hosts (see Table 3, where we also provide a summary of the
properties of the nuclei in the $H$ band, and of their optical-NIR or
NIR-NIR color properties).  Statistically, the distribution of $V-H$
colors is broader for the nuclei than for the surrounding galactic
structure, as seen in Figure 7 which compares the $V-H$ color
distribution for the nuclei with the $V-H$ color of the host galaxies.
This Figure contains only those galaxy-nucleus pairs for which an
isophotal fit to the underlying light of the host galaxies in both the
$V$ and the $H$ band could be performed, so that their $V-H$ color
could be derived by integrating the $V$ and $H$ radial light profiles
within 3$''$ (see Carollo \& Stiavelli 1998 and paper II for the $V$
and $H$ fits, respectively). The so-obtained colors are much less
affected by the effects of dust or recent star formation with respect
to aperture measurements. No isophotal fit could be performed for the
hosts of some of the nuclei of Figure 6, and in particular to the
hosts of the reddest and bluest nuclei; this explains the smaller
range in $V-H$ of Figure 7 with respect to Figure 6.  Even so, the
Kolmogorov-Smirnov, Kendall and Spearman tests give a probability of
0.996, 0.999 and 0.9997 that the two distributions are statistically
different.  This likely indicates ubiquitous presence of either
`activity' or dust reddening (or both) in the nuclei population.
Figure 8 a shows the $V$ absolute magnitude versus the $V-H$ color
for the nuclei; Figure 8 b shows their $J-H$ versus $V-H$ color-color
diagram. Different symbols in these Figures (are as in Figure 6 and)
are used to identify nuclei embedded in hosts belonging to a different
galaxy-class among the four discussed above.  A more detailed
discussion of these color-magnitude and color-color diagrams, and in
particular of the constraints that they put on the stellar population
properties of the nuclei, is given in Carollo et al.\ (2001); what we
remark here is that there is no obvious trends for any `segregation'
on these diagrams among nuclei embedded in hosts with different
circum-nuclear properties.

{\it (v)} Within the caveats imposed by the angular resolution of our
data and the distance of our targets, no significant difference is
found with Hubble type concerning the exact location of the nuclei
with respect to the underlying galactic structure, and their projected
shape, round versus elongated. In Table 3 we report, for each galaxy,
whether the compact nucleus is centered on, or offset from, the galaxy
isophotal center, and whether the nucleus and more generally the
nuclear region is round or elongated.  There are several cases where
our {\tt NICMOS} data suggest that the nuclei are not located
precisely at the isophotal centers of the host galaxies, as found
e.g. for M33 (de Vaucouleurs \& Freeman 1970; Minniti et al.\
1993). There are also a few cases where the nucleus shows some degree
of elongation, as also found for, e.g., the nucleus of M33 (Kormendy
\& McClure 1993; Lauer et al.\ 1998; Matthews et al.\ 1999).  With the 
current data, the displacement and elongation, however, appear not to
be correlated with either the luminosity and color of the nucleus, or
the galaxy type.  It remains an open issue whether some nuclei are
disk-like features, and whether the nuclei of galaxies with an AGN are
intrinsically different from those of non-AGN galaxies.  On the other
hand, the fact that many of the best-resolved nuclei are typically
round structures in projection supports the working hypothesis that
(at least in non-AGN galaxies and) for all Hubble types these nuclei
are typically cluster-like systems.

\section{Summary}

\noindent
We have presented the analysis of HST {\tt NICMOS} data in the F160W
($H$) filter for the central regions of a sample of 78 mostly
early-to-intermediate type spiral galaxies. For 58 of these systems we
have used {\tt WFPC2} F606W ($V$) images from our previous study to
investigate their $V-H$ nuclear colors. For 11 systems we have used
additional {\tt NICMOS} images in the F110W filter to investigate
their $J-H$ nuclear colors.  We have focused our attention on the
structural and color properties of the photometrically-distinct nuclei
that (similarly to what we had previously found in the visual band)
dominate the nuclear light in a large fraction of the sample in the
NIR, and on their galactic circum-nuclear regions.

Although the nuclear complexity that was unveiled in many of these
galaxies by the {\tt WFPC2} images is often still present in the $H$
band, the latter provides a cleaner view of the nuclear galactic
structure, and allows to see through areas of recent star formation
and through dust patches typically present in the central regions of
these system. This strengthens the findings which we had previously
suggested on the basis of our HST optical images, and adds further
clues for understanding the nature and origin of galactic nuclei. In
particular, several nuclear bar-like features have been detected in
the $H$-band, including some which have a $V-H$ or $J-H$ color
distinct from that of their surroundings.  Nuclear bars are however
not ubiquitous in this kind of systems.  Any relation between the
presence of a nucleus and of a barred component, if present, is not a
simple one, as nuclei are found in both barred and non-barred hosts.
In the $H$ band, consistent with what was found in the $V$ band, the
nuclei embedded in the bright early- to intermediate-type hosts span a
much larger magnitude range (about eight magnitudes) peak-to-peak than
the nuclei embedded in the fainter late-type disks.  In the $H$ band,
the brighter the nucleus the larger its radius, a correlation that was
found in the $V$ band and is confirmed by the NIR data.  Even
excluding the few bluer-than-average late-type nuclei in our sample,
the nuclei of early- to intermediate-type spirals span a large range
in $V-H$ colors, some being significantly redder (and very likely
dust-reddened), others being significantly bluer (`polluted' either by
a non-thermal component or by recent star formation), than their
surrounding circumnuclear regions. Several nuclei embedded in the
early- to intermediate-type hosts are offset (typically by a few tens
of pc) from the isophotal centers of the host galaxies; several are
flattened structures; however, most of the nuclei, at HST resolution,
are located at the galaxy centers, and appear to be round,
cluster-like structures. 

A quantitative analysis of the circum-nuclear galaxy structure and
strength of the nuclear stellar cusp slopes in the NIR is reported in
paper II.

\bigskip

\acknowledgements We thank Roeland van der Marel for providing the
pedestal-removal software, and the anonimous referee for his/her
careful review that has helped us improve the presentation of our
results.  This research has been partially funded by Grants
GO-06359.01-95A and GO-07331.02-96A awarded by STScI, and has made use
of the NASA/IPAC Extragalactic Database (NED) which is operated by the
Jet Propulsion Laboratory, Caltech, under contract with NASA.

\bigskip
\bigskip

\begin{center}{\bf Appendix A. NICMOS $H$ Band Imaging for Nine Galaxies}\end{center}

\noindent
In Figure 9 we present the {\tt NICMOS} F160W images of the nine
galaxies in our sample for which neither {\tt WFPC2} F606W nor F110W
images are available. The data reduction was done as described in \S
2. NGC1365 (a well-known AGN), NGC2903, NGC3031, NGC4569, and NGC6217
host photometrically-distinct nuclei; NGC 4569 also hosts a nuclear
disk. The presence of a nucleus is uncertain in NGC4536 and NGC4565;
no nucleus is present in NGC7479 and NGC7742.  Table 4 summarizes some
properties of the nine galaxies, including the $H$ measurements of
sizes and magnitudes for the nuclei.

\bigskip

\begin{center}{\bf Appendix B. Assessment of Systematic Effects}\end{center}

\noindent
Measuring the half-light radius and luminosity of a distinct nucleus
embedded in a galaxy core entails assumptions on how to partition the
observed light between the underlying galaxy and the nucleus. The
different assumptions that we have used in \S 2.2, while all yielding acceptable
decompositions, provide however different results, since it is not
easy to establish where the nucleus ends and the galaxy begins. The
situation is therefore one where the systematic error arising from
separating the galaxy light from the nucleus light outweighs the error
on the measurement of the radius and the luminosity of the nucleus
itself. To assess the systematic errors on these quantities, we
performed two sets of tests. The results for the F160W filter are
described below. A similar analysis was performed for the F110W and
the {\tt WFPC2} F606W filters, reaching similar conclusion as for the
F160W filter.

First, we tested the sensitivity of our nucleus-extraction procedure
in the region of parameter space covered by those nuclei embedded in
hosts for which an isophotal fit could be performed, and whose nuclear
surface brightness profiles could be modeled with a Nuker-law (Lauer
et al.\ 1995; see paper II for the description of our isophotal and
Nuker fits).  The Nuker law reduces to a power law when the
characteristic radius $r_b$ is much larger than the nucleus' radius.
We built model galaxy images starting from the best-fit analytical
descriptions of the light underlying the nuclei in these galaxies, and
added to each of these model galaxies a contribution from a nucleus of
size and luminosity equal to those measured with our procedure in that
specific system.  Two different analytical forms were adopted to
describe the nuclei, a ``generalized'' modified-Hubble law, which
reads:
\begin{equation}
\label{hubblelaw}
I(R) = \frac{L}{2 \pi b^2} \left(1+\frac{R^2}{b^2} \right)^{-5/2}
\end{equation}

\noindent
and a Plummer law, which reads:

\begin{equation}
\label{plummerlaw}
I(R) = \frac{L}{b^2} \left(1+\frac{R^2}{b^2} \right)^{-2}.
\end{equation}

\noindent
In both Eq.\ (1) and (2) $b$ is the scalelength and $L$ is the total
luminosity. We then convolved the model galaxy-plus-nucleus images
with the {\tt NICMOS} PSF in the appropriate filter (obtained with
TINYTIM, as done for the data), and applied our nucleus-extraction
procedure to these PSF-convolved model-images.  The radius was
corrected for PSF broadening in the same way as was done for the
actual data.  This test provides an estimate of how accurately we can
disentangle the nuclei from their hosts for galaxy-plus-nucleus pairs
of the kind that we have detected (similar sizes/luminosities of
the clusters; similar contrasts with the underlying/surrounding
galaxy).

Figures 10 a and b show, for these galaxy-nucleus pairs, the
distribution of the ratio between the recovered flux and the input
flux, and the distribution of the ratio between the recovered radius
and the input radius, respectively.  Figure 10 c shows $\Delta mag =
(m_{recover}-m_{input})$,  the difference between the recovered
and the input magnitude, as a function of the logarithm of the ratio
between recovered and input radius (after correction for systematic
shifts).  The figures show that 90\% of the measurements lie within
$\approx$0.3 magnitudes from the expected values, and that our extraction
procedure applied to the intrinsic nuclear light profiles
systematically underestimates the flux on average by $\approx$0.4
magnitudes, with a scatter of $\approx$0.2 magnitudes.  By contrast, the
mean radius appears to be overestimated in average by $\approx$45\%, with
a scatter of $\approx$25\%; the 50\%-error includes $\approx$80\% of the
points.  A weak correlation between the derived radius and magnitude
of the nuclei is detected, but the maximum absolute variations are
rather modest, about a factor 3 increase in radius  for a $\approx$1
magnitude brightening of the nucleus.

Secondly, we performed another set of tests which covered a larger
area of parameter space to assess the degree of degeneracy in the
parameters provided by our galaxy-nucleus decomposition. These
additional tests allow us to estimate how much and often
intrinsically different galaxy-nucleus pairs can result in similar
measurements when analysed with our procedure. We built a grid of
model galaxies described by either a power-law cusp of slope $\gamma$
ranging from 0 to 0.7, or an exponential profile, and we added to
these model galaxies the contribution from a central nucleus,
described by a Plummer law. The nucleus was assumed to have
$H_{AB}=19.6$; the surface brightness of the underlying galaxy was
taken similar to the typical one observed for real objects. In
particular, a mean $H_{AB}$ surface brightness within $20''$ equal to
19.7 mag/arcsec$^2$ was considered for the power-law galaxies,
corresponding to a central surface brightness after convolution with
the PSF of 19.2, 18.4, 17.4 or 16.4 mag/arcsec$^2$ for $\gamma = 0.1$,
0.3, 0.5 or 0.7, respectively.  For the two highest $\gamma$ values of
0.5 and 0.7, also a mean surface brightness within $20''$ equal to
20.5 mag/arcsec$^2$ was considered, corresponding to a central surface
brightness after convolution with the PSF of 18.2 or 17.2
mag/arcsec$^2$, respectively.  A central surface brightness equal to
18.5 or 19.3 mag/arcsec$^2$ was considered instead for the exponential
galaxies.  We convolved this set of galaxy-nucleus models with the
appropriate PSF, and applied our procedure to recover the input
parameters.

Figure 11 a shows the $\Delta mag$ (the difference between the
recovered and input $H_{AB}$ magnitude of the nucleus) versus the
nuclear cusp slope $\gamma$ for the power-law galaxies, and versus the
exponential scale-length for the exponential galaxies.  Figure 11 b
shows the recovered- versus the input-FWHM of the central nucleus.
Notice how difficult it is to recover the radius of the nucleus when a
large nucleus lies on top of an underlying steep cusp.  Figure 11 c
shows the $\Delta mag$,  the difference between the recovered and
the input magnitude, versus the recovered radius of the nucleus.  The
results of this set of tests demonstrate that when all the
galaxy-nucleus pair models are taken into account, the flux is
underestimated by $\approx$0.27 magnitudes, with a scatter of 0.60
magnitudes; 80\% of the points lie within the 50\% flux error.  The
radius appears to be overestimated by $\approx$30\% with an rms of
$\approx$40\% and with the 50\%-error including 90\% of the points.  Also
these general models - especially the power-law ones - show a weak
trend between derived radius and magnitude of the nuclei. The observed
maximum variation is again about a factor 3 of increase in radius,
which in this case corresponds to $\approx 1.5$ magnitude brightening
of the nucleus.

Summarizing, both the tests performed on specific galaxy-nucleus pairs
and the more general ones indicate that: (1) the reported fluxes are
likely underestimated by $\approx$0.3 magnitudes, and are accurate to
better than 0.6 magnitudes; (2) the reported radii are likely
overestimated by about 40\% and are accurate to within 40\%; (3) the
galaxy-nucleus degeneracy and possibly the adopted extraction
technique may produce a mild trend for brighter nuclei to be
slightly more extended than fainter nuclei, although it is unlikely
that such an artifact could produce a correlation between luminosity
and size of the nuclei involving several magnitudes of difference in
luminosity. 

We finally checked whether there was any major systematic effect
affecting our measurements depending on the luminosity of the nucleus
embedded in a given galaxy. We computed an additional set of about 200
galaxy-models with a central surface brightness of 18.4 mag
arcsec$^{-2}$ and nuclei with a range of luminosities; we used both
power-laws and exponentials models with cusps and exponential lengths
covering the same intervals as in the models of Figure 10.  While the
systematic error in the measurement increases, there is no systematic
trend on the recovered luminosity of the nucleus. This is illustrated
in Figures 12 a and b where we plot, as a function of the input
magnitude of the nucleus, the difference between the recovered and the
input magnitudes for both power-law and exponential galaxy-models,
respectively.  For each luminosity bin the plot shows the mean shift
and the error; the latter is dominated by systematic effects.

Although these experiments do provide an estimate of the size of our
systematic errors, we have chosen not to apply these ``corrections''
to the radii and luminosities of the nuclei in our analysis, since:
{\it (i)} the absolute amount of the systematic shifts is not such
to affect in any significant way our conclusions, and {\it (ii)} the
estimated correction would be strictly valid if all nuclei were
``homologous'', i.e., had similar intrinsic profiles, and could instead
introduce additional scatter if the intrinsic light profile changed
from nucleus to nucleus.

 The combined results for the {\tt WFPC2} and {\tt NICMOS} tests were
used to assess the uncertainties associated with the colors of the
nuclei.  On the assumption that nuclei and underlying galaxies have
similar intrinsic light profiles in the visual and NIR and do not have
large unresolved color gradients, the same analytical models would
properly describe both the $V$ and the $H$ band data. In this case one
expects the error on the $V-H$ color of the nuclei to be much smaller
than the total formal error obtained by quadratically adding the
errors on each band. We checked this assumption on the galaxy-nucleus
pairs of Figures 10 a and 10 b. In Figure 13 we show the histogram of the
error on the $V-H$ color obtained adopting for the nuclei a modified
Hubble profile (solid line) or a Plummer law (dotted line): The figure
shows that although the systematic effect is not entirely cancelled,
the resulting error is reduced to 0.4 magnitudes instead of the formal
0.8 magnitudes.

\newpage

\begin{figure}
\caption{Panel a: $H$ images (left) and $V-H$ color maps (right) for
38 (out of 39, see text) galaxies observed with {\tt NICMOS} which
host a distinct nucleus and for which also {\tt WFPC2} images are
available from our previous F606W survey.  Panel b: $H$ images (left)
and $J-H$ color maps (right) for the 5 galaxies observed in F160W
which host a distinct nucleus and for which the only additional image
available from our program is in the F110W filter. All images have
North up and East left, and are $9'' \times 9''$ size.}
\end{figure}

\begin{figure}
\caption{Same as Figure 1, but for the 25 galaxies which either do not
host a distinct nucleus or in which the presence of a distinct nucleus
is uncertain (19 objects with $V-H$ color maps, panel a,  and 6 objects with
$J-H$ color maps, panel b).}
\end{figure}

\begin{figure}
\caption{Panel a (left): $H$ light profile of NGC 2082. The galaxy
hosts a distinct nucleus, clearly visible on top of the underlying
exponential bulge. Panel b (right): $H$ light profile of NGC 2460. The
galaxy hosts an $R^{1/4}$-law bulge and no detectable distinct
nucleus.}
\end{figure}

\begin{figure}
\caption{A comparison of the sizes of the nuclei as determined from
the F160W images and the F606W (triangles) or F110W (squares) images.}
\end{figure}

\begin{figure}
\caption{Panel a (left): The absolute magnitude versus half-light
radius relation for the nuclei in the F160W passband. Panel b (right):
For comparison, the same relation is plotted for the {\tt WFPC2} F606W
passband.  Triangles represent nuclei embedded in early- to
intermediate-type hosts, S0a to Sc; filled squares are the four nuclei
embedded in the Scd and later-type disks.  }
\end{figure}

\begin{figure}
\caption{The $V-H$ color of the nuclei versus the RC3 Hubble type of
the host galaxy. Different symbols identify the four `classes' of
galaxies described in \S 3. In particular, filled triangles are the
galaxies with concentrated nuclear star formation mixed with dust
(class 1), open squares are the galaxies with diffuse blue nuclear
regions (class 2), pentagons are the galaxies with regular
nuclear/circum-nuclear dust (class 3), and 5-points stars are the
galaxies with irregular nuclear/circum-nuclear dust (class 4).}
\end{figure}

\begin{figure}
\caption{The $V-H$ color distribution for the nuclei (solid line) and
the underlying host galaxy (dotted line). The latter colors are
obtained integrating within 3$''$ the Nuker fits to the radial light
profiles (profiles and fits presented in paper II).}
\end{figure}

\begin{figure}
\caption{Panel a: Absolute $V$ magnitude versus $V-H$ color for the nuclei.
Panel b: $J-H$ versus $V-H$ color for the nuclei. Symbols are as in
Figure 6.}
\end{figure}

\begin{figure}
\caption{The $H$ images for the nine galaxies observed only in this
filter. }
\end{figure}

\begin{figure}
\caption{Results of the tests performed for the subset of
`really-observed' galaxy-nucleus pairs to assess the accuracy of the
recovered nucleus-parameters. Panel a: The distribution of the ratio
between the recovered radius and the input radius of the simulated
nucleus. Panel b: The distribution of the ratio between recovered- and
input-flux of the nucleus. In Panels a and b the dashed line refers to
modified-Hubble nuclei, the dotted line to Plummer nuclei, and the
solid line to the sum of the two families.  Panel c: The difference
between input and recovered radius as a function of the difference
between input and recovered luminosity of the nucleus, respectively.
Empty symbols refer to modified-Hubble nuclei, filled symbols to
Plummer nuclei.}
\end{figure}

\begin{figure}
\caption{Results of the general tests performed on a grid of power-law
and exponential galaxies to assess the accuracy of the recovered
parameters for the nuclei. Panel a: The difference between the
recovered and input $H_{AB}$ magnitude of the nucleus ($\Delta mag$)
versus the nuclear cusp slope $\gamma$ for the power-law galaxies
(left) and versus the exponential scale-length for the exponential
galaxies (right).  For both the power-law and the exponential
galaxies, squares and triangles identify the brightest and the
faintest sets of simulated host galaxies, respectively; the size of
the symbols increases with increasing input scale length of the
Plummer nucleus (1, 1.5, 2, 3, 4 and 5 pixels, respectively). Panel b:
The recovered- versus the input-FWHM of the central nucleus for the
power-law (left) and the exponential galaxies (right).  Squares and
triangles identify again the brightest and the faintest sets of
simulated host galaxies. For the power-law galaxies, the size of the
symbols increases for increasing $\gamma$ (from 0 to 0.7 for the
bright hosts, and from 0.5 to 0.7 for the faint hosts).  For the
exponential galaxies, the size of the symbols increases with
increasing scale-length $h$ of the exponential-type host ($h = 5, 10,
25$ pixels). For reference, the solid line represents the effect of
the PSF on an `isolated' nucleus (in the absence of underlying
galaxy).  Panel c: The $\Delta mag$ versus the input radius of the
nucleus for the Nuker-law (left) and the exponential galaxies
(right). For the power-law galaxies, symbols are crosses, empty
triangles, empty squares and empty pentagons for the $\gamma=0.1, 0.3,
0.5, 0.7$ bright hosts, respectively; filled squares and pentagons
represent instead the $\gamma=0.5,0.7$ faint hosts.  The size of the
symbol increases with increasing input scale length of the Plummer
nucleus (again 1, 1.5, 2, 3, 4 and 5 pixels, respectively).  For the
exponential galaxies, empty and filled symbols represent the bright
and faint families of hosts, respectively; triangles, squares and
pentagons represent respectively the $h = 5, 10, 25$ cases; the size
of the symbols increases with increasing input scale length of the
nucleus.  The solid line represents the effect of the PSF on an
isolated nucleus.}
\end{figure}

\begin{figure}
\caption{Panel a (left): The difference between the recovered and the
input magnitudes of a nucleus as a function of the input magnitude of
the nucleus for the power-law galaxy models. Panel b (right): The
similar plot for the exponential galaxy-models. The plots show that
there is no systematic trend with luminosity, although the systematic
errors increase with decreasing luminosity.}
\end{figure}

\begin{figure}
\caption{Results of the tests performed for the subset of
`really-observed' galaxy-nucleus pairs of Figure 9 to assess the error on the V-H
colors. The histograms refer to nuclei modelled with either a modified
Hubble profile (solid line), or a Plummer law (dotted line). The sum
of the two histograms is a conservative benchmark for determining the
error, given that the intrinsic shape of the nuclei is unknown. The
partial cancellation of systematic effects, which are similar in both
bands, leads to an error on the $V-H$ color of $\sim0.4$ magnitudes.}
\end{figure}

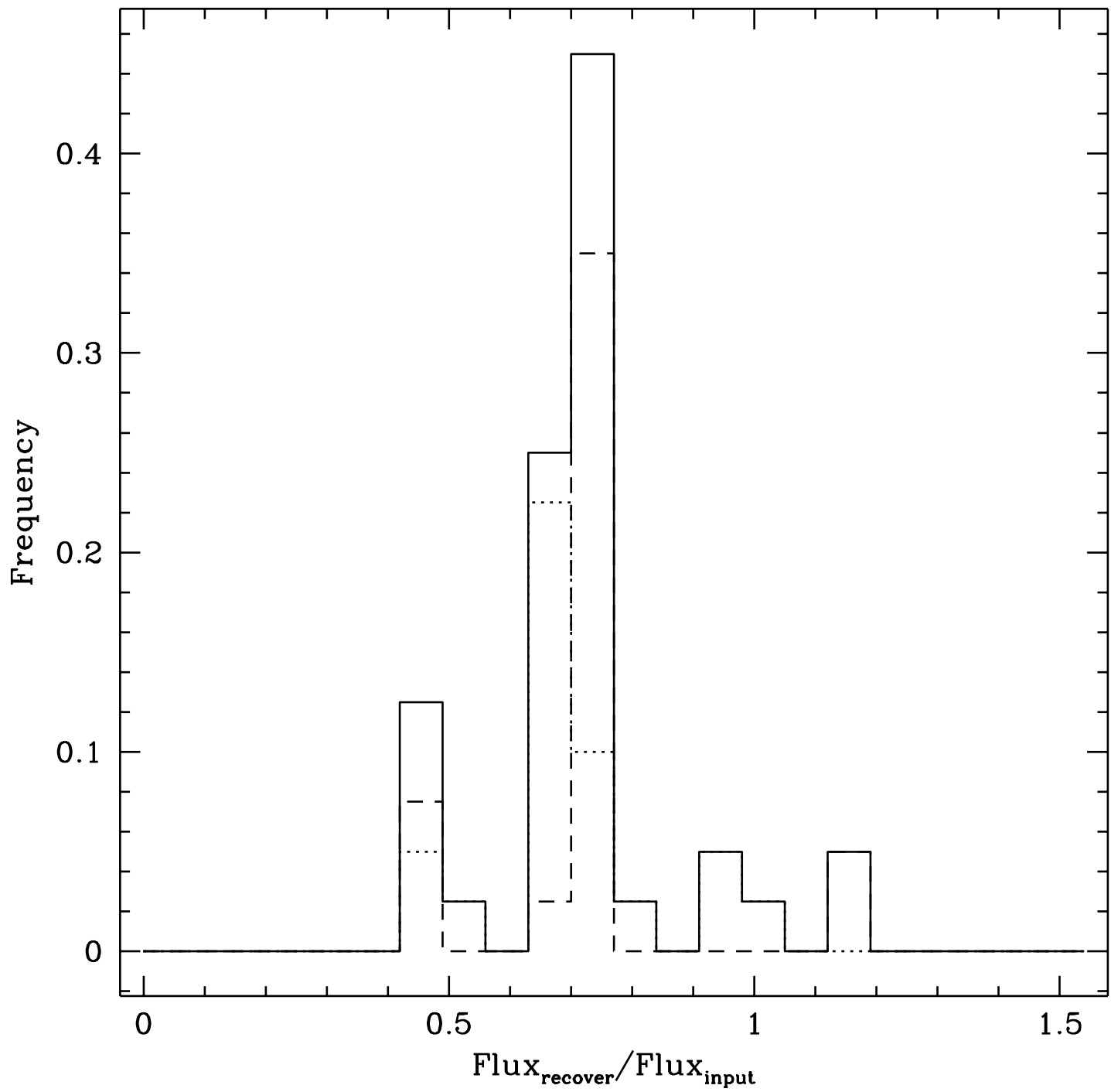
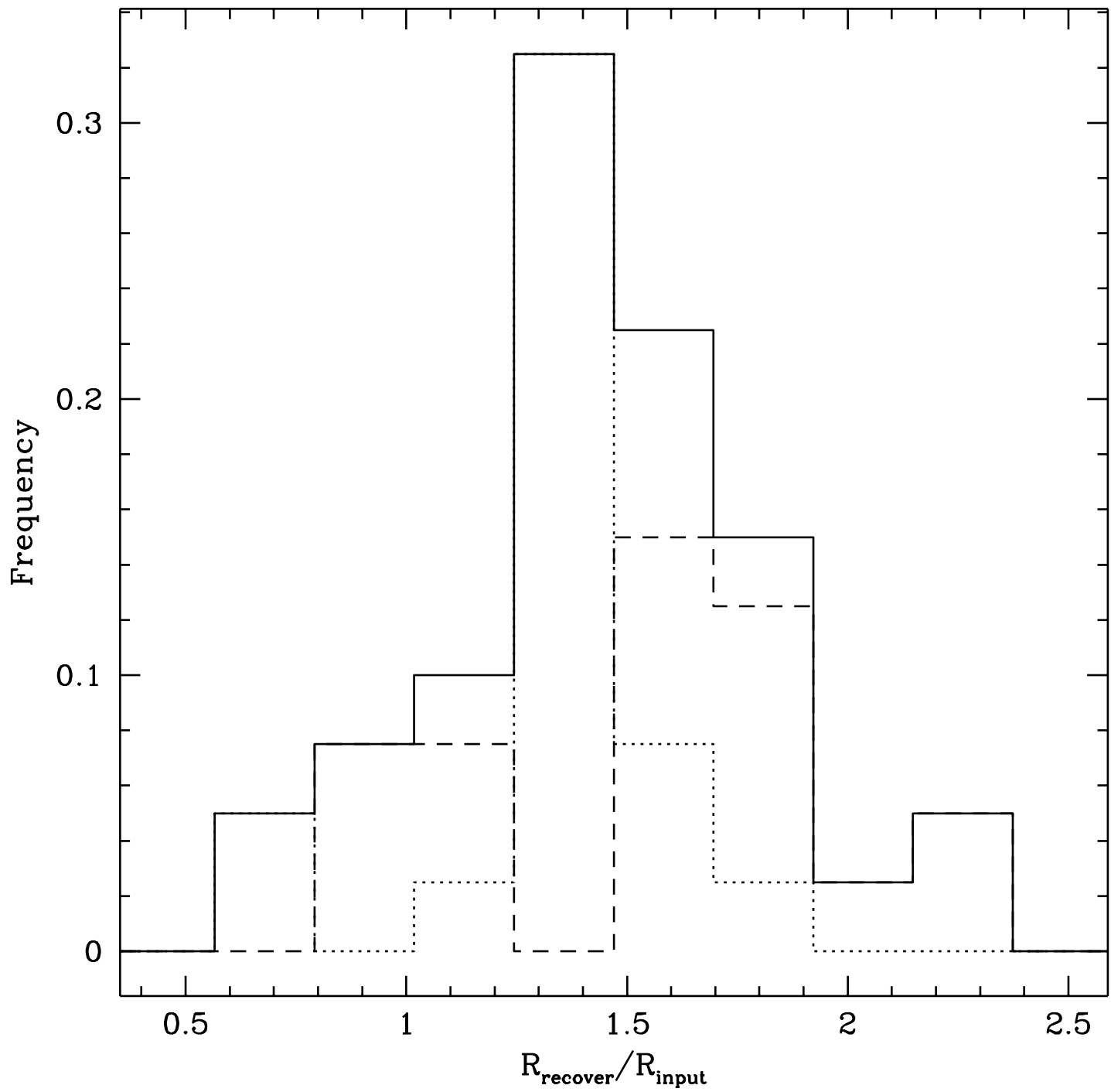
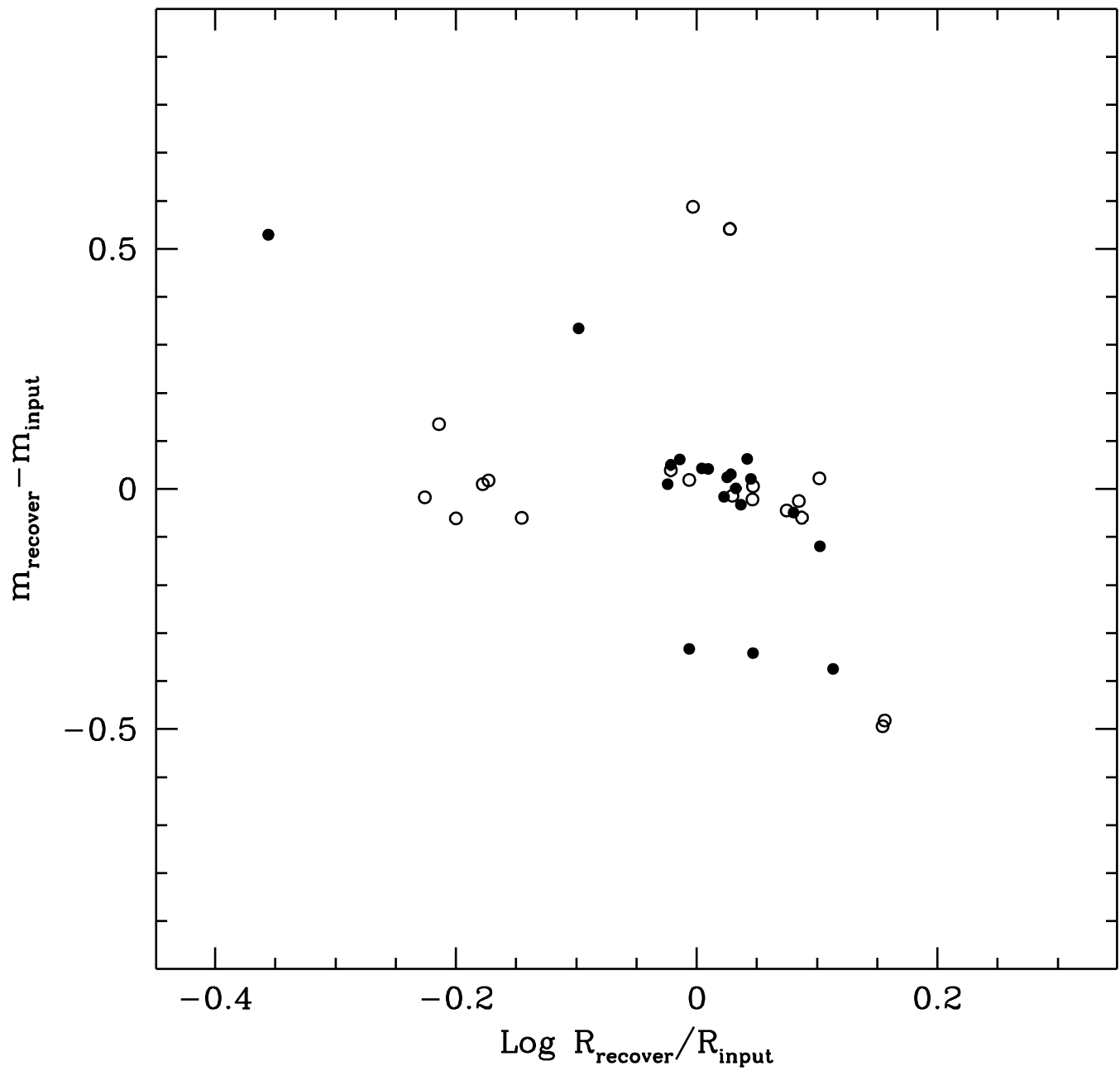
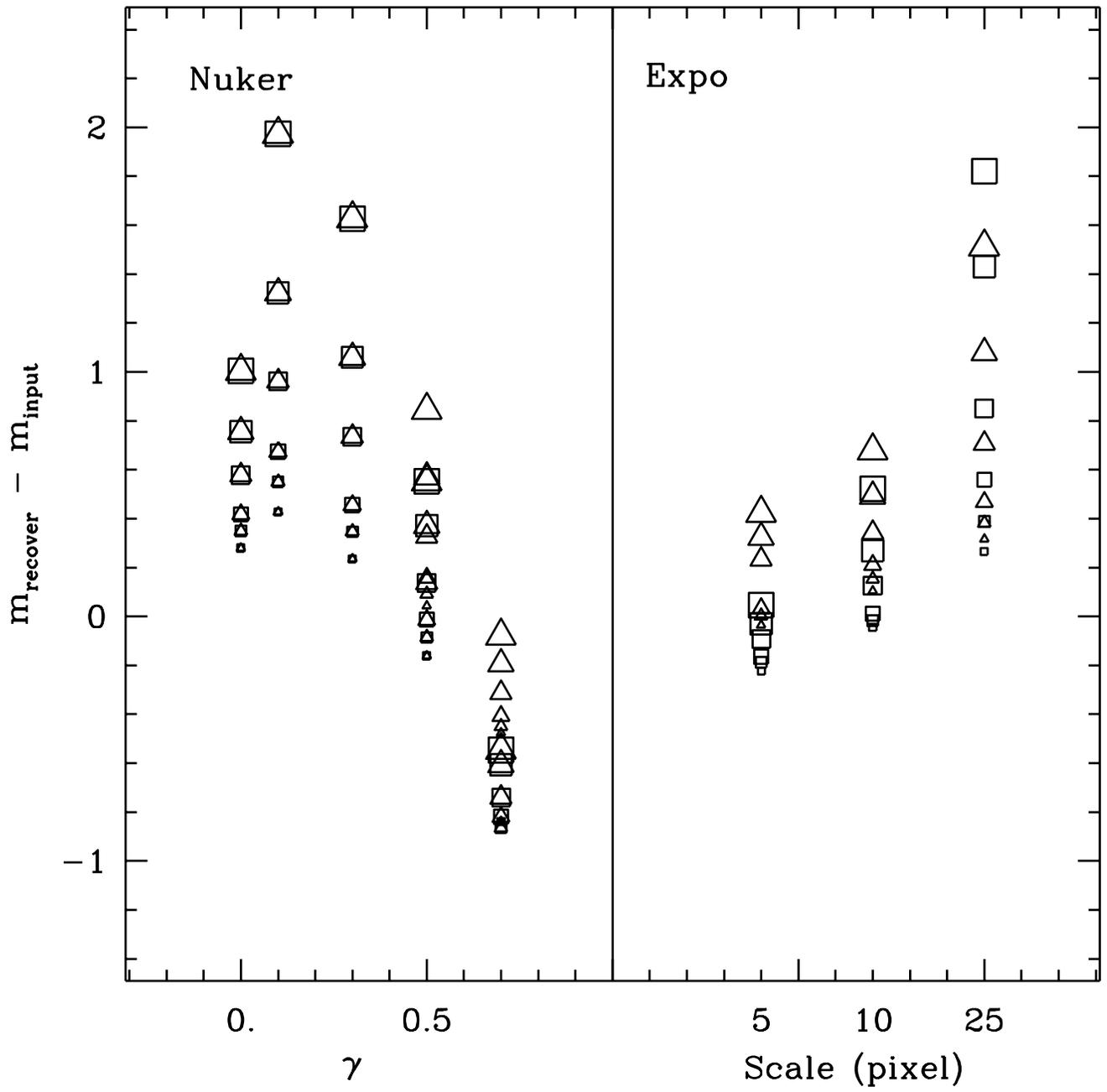
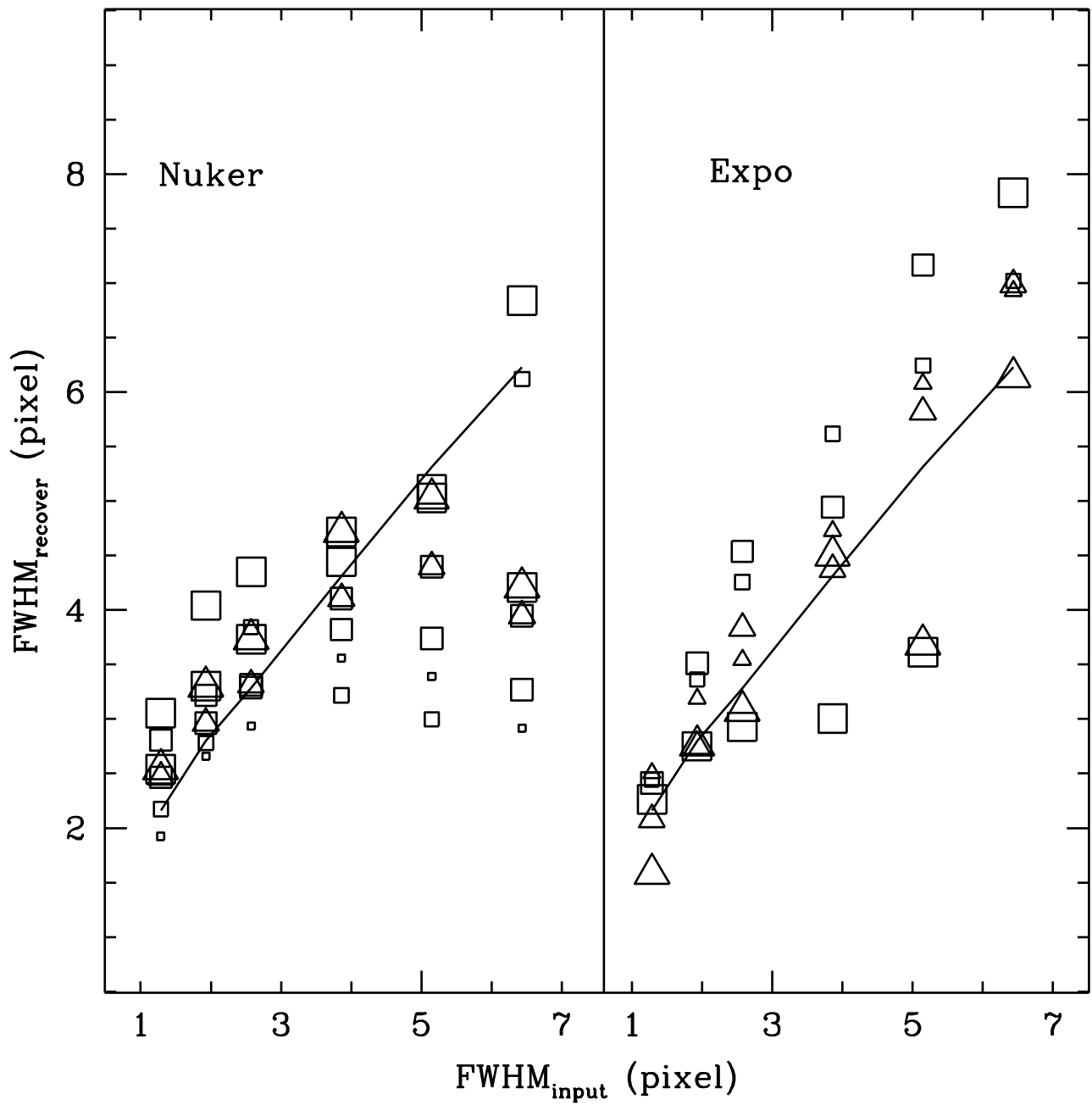
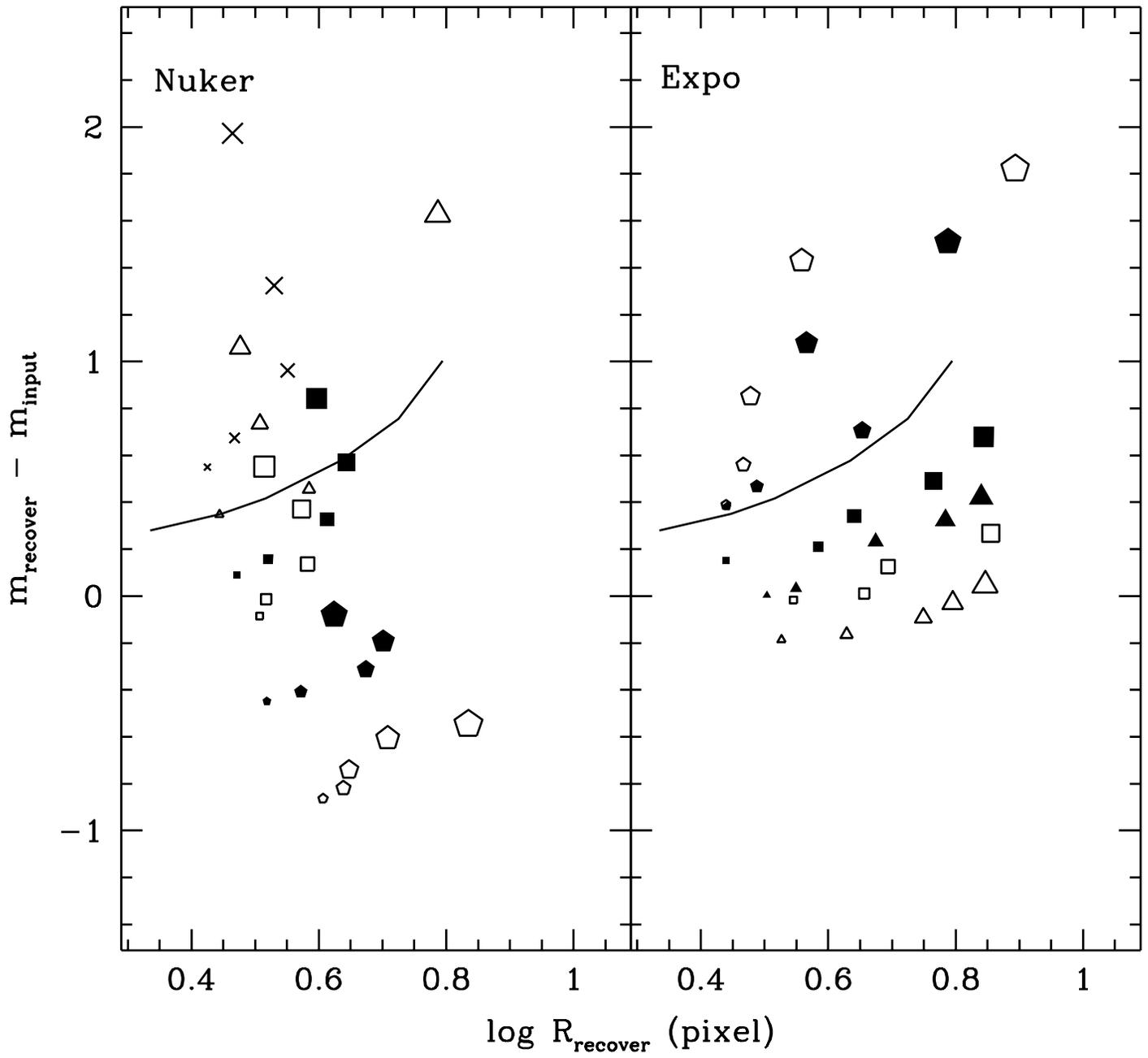
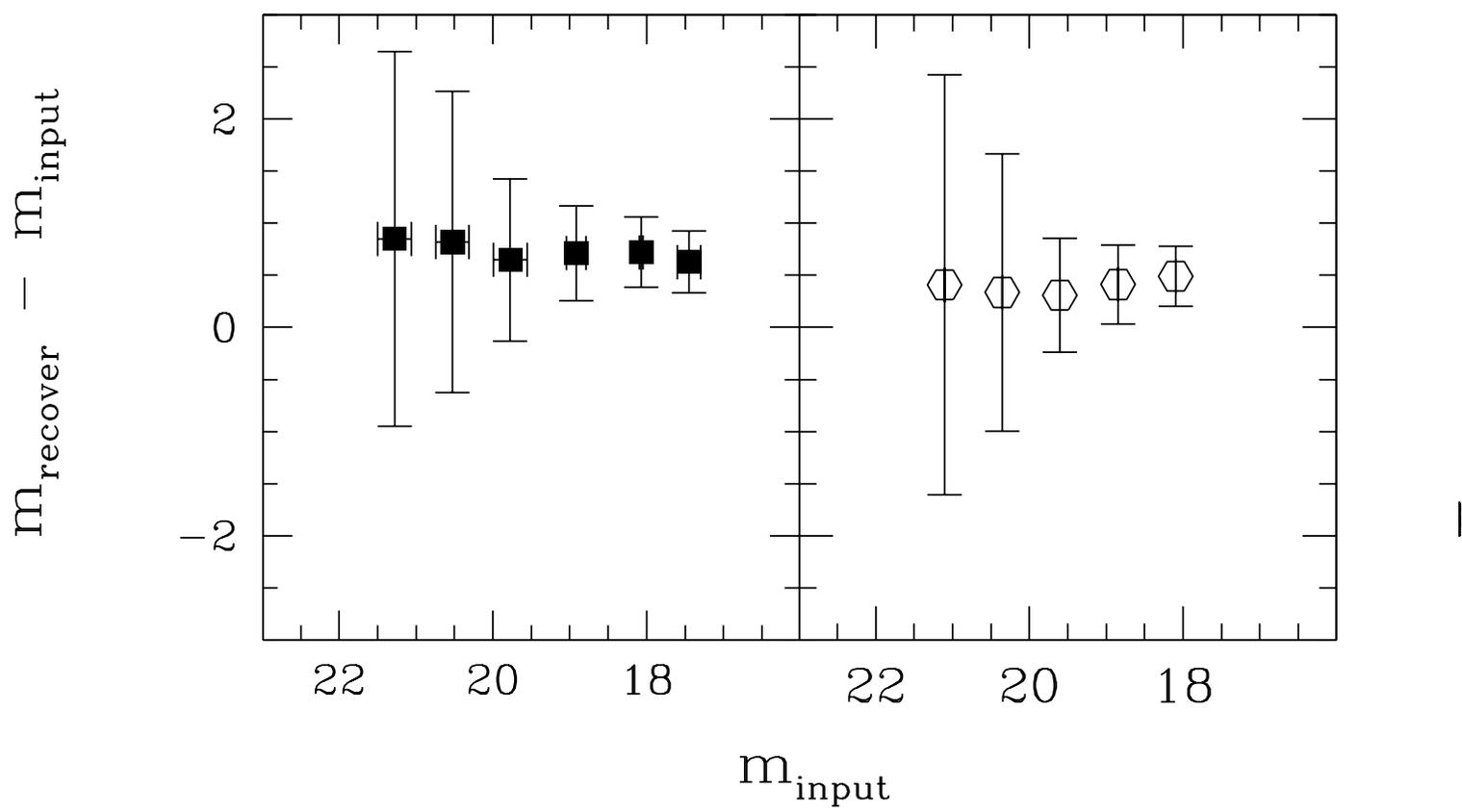
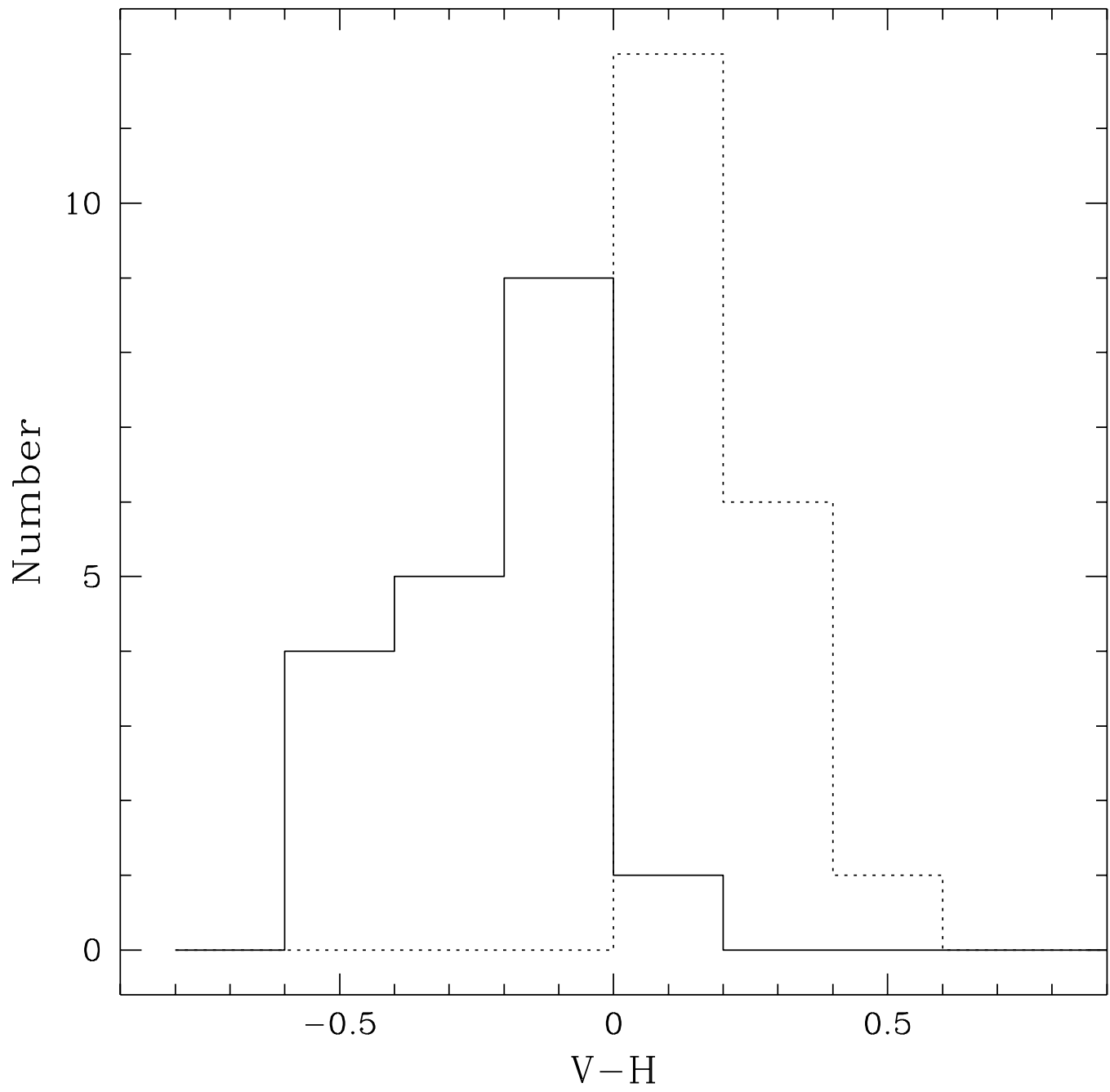
\end{document}


\begin{table*}
{\tiny\begin{center}\begin{tabular}{lccccccc}
\hline\hline
\multicolumn{1}{l}{Name} &
\multicolumn{1}{c}{Filter} &
\multicolumn{1}{c}{$D$} &
\multicolumn{1}{c}{Type} &
\multicolumn{1}{c}{$m_B$} &
\multicolumn{1}{c}{$A_B$} &
\multicolumn{1}{c}{Bulge} & 
\multicolumn{1}{c}{Spectrum} \\
&&(Mpc)&&(mag)&(mag)&&\\
\hline \tiny
E205G7	&F606W, F110W&   31 		&.SBR3..&15.02$^*$& 0.24    &-& HII \\
E240G12	&F606W, F110W&   28 		&.S?....&14.35$^*$&	0.0  &EXPO   & HII  \\  
E404G3	&F606W$^*$, F110W& 	37 	&.SBT4..&14.04$^*$& 0.0     &-& - \\
E498G5	&F606W&   	37 		&.SXS4P.&13.54    &	0.42    &EXPO& HII \\  
E499G37	&F606W&  	 14 		&.SXS7..&13.15$^*$&	0.17    &EXPO& -   \\  
E548G29	&F606W&  	 24 		&.SB?...&14.19$^*$&	0.04    &EXPO& - \\ 
E549G18	&F606W, F110W &  24 		&.SXT5..&13.49$^*$&	0.11    &EXPO& - \\ 
E572G22	&F606W, F110W&   22 		&RSB.7P?&14.92$^*$&	0.08   & EXPO& - \\
IC4390	&F606W&  	 30 		&.SXS4..&13.41$^*$&	0.40  & - & HII \\  
IC5271	&F110W &    	27 		&.S..3?.&12.34&	0.04    &-& - \\ 
IC5273	&F606W$^*$, F110W&    21 	&.SBT6..&11.80&	0.04    &-& - \\ 
N289	&F606W$^*$, F110W&    20 	&.SBT4..&11.70&	0.02    &-& - \\
N406	&F606W, F110W&   23 		&.SAS5..&13.04&	0.05    &EXPO& - \\
N488	&F606W&  	 37 		&.SAR3..&11.06&	0.09    &$R^{1/4}$& INT  \\  
N772	&F110W&    	38 		&.SAS3..&10.92&	0.17    &-& INT\\  
N972	&F606W$^*$, F110W&    24 	&.S..2..&11.90&	0.37    &-& HII  \\
N986	&F606W&  	31 		&.SBT2..&11.61&	0.03    &-& HII  \\
N1345	&F606W&  	 24 		&.SBS5P.&14.23&	0.10    &EXPO& - \\ 
N1398	&F110W&    	20 		&PSBR2..&10.57&	0.0     &-& - \\
N1483	&F606W&  	 16 		&.SBS4..&13.11 &	0.0  & EXPO  & - \\ 
N1688   &F606W&  	 17 		&.SBT7..&12.00&	0.0     &-& - \\  
N1800	&F110W&    	12 		&.IBS9..&13.13&	0.0   &- &- \\
N1892	&F606W&  	 17 		&.S..6..&12.70&	0.13    &-& - \\ 
N2082	&F606W, F110W&   19 		&.SBR3..&12.47&	0.15    &EXPO& - \\ 
N2196	&F606W, F110W&   35 		&PSAS1..&11.37&	0.45    &$R^{1/4}$& - \\
N2339	&F606W, F110W&    36 		&.SXT4..&11.90&	0.61    &-& INT  \\  
N2344	&F606W&  	 14 		&.SAT5..&12.39&	0.42    &$R^{1/4}$& - \\ 
N2460	&F606W&  	 26 		&.SAS1..&12.58&	0.14    &$R^{1/4}$& - \\  
N2566	&F110W&    	25 		&PSBT2P.&10.35&	1.48    &-& - \\
N2748	&F606W&  	 24 		&.SA.4..&12.40&	0.01    &-& HII  \\  
N2758	&F606W&  	 30 		&PSB.4P?&13.46$^*$&	0.53 &EXPO   & - \\  
N2964	&F606W&  	 20 		&.SXR4..&11.97&	0.02    &- &HII  \\  
N3067   &F606W, F110W &       23 	&.SAS2..&12.71& 0.07    & -&HII	 \\  
N3259   &F606W&       27 		&.SXT4..&12.89& 0.0     &EXPO & -    \\
N3277   &F606W, F110W &       21 	&.SAR2..&12.47& 0.03    &- &HII   \\  
N3455   &F606W, F110W&       19 	&PSXT3..&12.83$^*$& 0.02   &EXPO &  -    \\  
N3898   &F606W&       22 		&.SAS2..&11.60& 0.0     & $R^{1/4}$ &INT \\	      
N3900     &F606W&       27     		&.LAR+..&12.13& 0.07     & $R^{1/4}$& AGN      \\
N3928     &F606W&       17     		&.E?....&13.18& 0.04     & -& HII      \\
N3949       &F110W&       12	  	&.SAS4..&11.51& 0.03     & -& HII      \\
N4102     &F606W&       13     		&.SXS3..&11.98& 0.01     & -& AGN      \\
N4219     &F606W&       29     		&.SAS4..&12.10& 0.61     & -& -     \\
N4384     &F606W&       37     		&.S..1..&13.50& 0.0      & -& HII      \\ 
N4527     & F110W&       34     	&.SXS4..&11.36& 0.02     & -& INT    \\
N4750     &F606W, F110W&       27     	&RSAT2..&12.04& 0.03     & -& AGN       \\
N4806     &F606W, F110W&       37     	&.SBS5?.&13.09$^*$& 0.35     &-&  -      \\
N4980     &F606W&       22     		&.SXT1P?&13.19$^*$& 0.34    & EXPO&  -      \\
N5188 &F606W, F110W&       36     	&PSXS3P.&12.79& 0.17     &- & HII       \\
N5377     &F606W, F110W&       22     	&RSBS1..&12.24& 0.0      &-&  AGN       \\
N5448 &F606W &          32 		&RSXR1..&11.89 & 0.04 & - & AGN\\
N5678     &F606W$^*$, F110W&   30  	&.SXT3..&12.11& 0.02     &- & INT       \\
N5806     &F606W, F110W&       29     	&.SXS3..&12.23& 0.17     & -& INT \\
N5985     &F606W, F110W&       46     	&.SXR3..&11.85& 0.02     & $R^{1/4}$& AGN       \\
N6000     &F606W&       32     		&.SBS4..&12.30$^*$& 0.72   & - &  HII    \\
N6239     &F606W, F110W&       14     	&.SBS3P&12.93& 0.01     & -& HII       \\
N6340     &F606W&       18     		&.SAS0..&11.66& 0.21     &$R^{1/4}$&  AGN       \\
N6384     &F606W, F110W&       26     	&.SXR4..&10.70& 0.44     &EXPO&  INT       \\
N6810     &F606W, F110W&       30     	&.SAS2.&12.19& 0.18     &- & INT \\
N6951     &F110W&       22  		&.SXT4..&11.68& 0.94     &- & AGN       \\
N7013     &F606W&       13     		&.SAR0..&11.10& 1.30     & -& AGN       \\
N7162     &F606W$^*$, F110W&   33 	&.SAS5..&13.29& 0.0      & -& -      \\
N7177     &F110W&       18  		&.SXR3..&12.01& 0.25     & -& INT       \\
N7188     &F606W$^*$, F110W&       27  	&PSBT4..&13.95*& 0.07     &-&  -      \\
N7217     &F110W&       15   		&RSAR2..&11.02& 0.44     &-&  AGN       \\
N7259     &F606W$^*$, F110W&   26  	&.S..3..&13.88$^*$& 0.0   &-   &  -      \\
N7280     &F606W&       28     		&.LXR+..&12.78& 0.23     &$R^{1/4}$&  AGN       \\
N7421     &F606W, F110W&       26     	&.SBT4..&12.48& 0.04     &-&  Q   \\
N7513     &F110W&       24  		&PSBS3P.&11.96& 0.19     &-&  -      \\
N7690     &F606W, F110W&       21     	&.SAR3..&13.00& 0.0      &-&  HII \\
\hline
\end{tabular}\end{center}}
\caption{The 69 galaxies of our sample observed with {\tt NICMOS}
Camera-2 in F160W. The column `Filter' lists the additional passbands
in which these galaxies have also been observed in our program (an
asterisk close to the F606W filter identifies the 8 galaxies for which
the analysis of the {\tt WFPC2} photometry is not included in our
previous papers, due to their delayed observations). The remaining
columns give the distance in Mpc, the morphological type and the
apparent total B magnitude from the RC3 (asterisks indicate Cousins B measurements), and the $B$ extinction from
Burstein \& Heiles (1984).  The column `Bulge' reports whether
the galaxies host either
 an $R^{1/4}$ or an exponential bulge, based on the analysis 
of our {\tt WFPC2} images (the radial extent of the {\tt NICMOS} data
is insufficient to properly perform a bulge/disk decomposition). 
When available, in the last column we
report the information on the central spectral properties shown by the
system at ground-based resolution (HII = HII-type; AGN = AGN-type, INT
= Intermediate between HII and AGN; Q = Quiescent; information
extracted from Ho et al.\ 1997, and from our own unpublished spectra,
Carollo, Chiappini \&  Heckman 2001, in preparation).}
\label{tab1}
\end{table*}
\normalsize

\newpage 
\begin{table*}
{\tiny\begin{center}\begin{tabular}{lcccccc}
\hline\hline
\multicolumn{1}{l}{Name}  &
\multicolumn{1}{c}{$m_V$} &
\multicolumn{1}{c}{$R_V$} &
\multicolumn{1}{c}{$m_J$} &
\multicolumn{1}{c}{$R_J$} &
\multicolumn{1}{c}{$m_H$} &
\multicolumn{1}{c}{$R_H$} \\
&(mag)&($''$)&(mag)&($''$)&(mag)&($''$)\\
\hline \tiny
E205G7  &  19.2(0.2)  &  0.05(0.02) &  17.7(0.6)  &  0.12(0.02) &  17.3(0.8)  &  0.12(0.02) \\
E240G12 &  22.2(0.1)  &  0.02(0.02)$\dagger\dagger$       &  21.4(0.5)  &  0.02(0.02) &  21.2(0.6)  &  0.03(0.02)\\ 
E404G3$^*$  &   22.1(0.4)      &    0.054(0.08)   &  20.9(0.3)  &  0.08(0.01) &  20.8(0.9)  &  0.03(0.02)\\ 
E498G5  &  20.0(0.1)  &  0.04(0.02) &  -      &    -    &  18.7(0.3)  &  0.05(0.02)\\
E499G37 &  20.6(0.7)  &  0.36(0.05) &   -     &   -     &  20.3(0.2)  &  0.28(0.06)\\	 
E548G29 &  21.8(0.4)  &  0.08(0.03) &  -      &    -    &  20.6(0.6)  &  0.04(0.01)\\	 
E549G18 &  23.3(0.1)  &  0.09(0.01) &  20.7(0.3)  &  0.10(0.01) &  20.3(0.7)  &  0.17(0.07)\\ 	 
E572G22 &  21.8(0.1)  &  0.06(0.02) &  21.2(0.2)  &  0.04(0.01) &  21.1(0.4)  &  0.04(0.01)\\
IC4390  &  18.1(0.3)  &  0.26(0.09)       &   -     &    -    &  16.4(0.3)  &  0.17(0.08)\\
IC5271  &   -     &    -    &  16.8(0.1)  &  0.13(0.02) &  16.4(0.2)  &  0.11(0.03)\\	 
IC5273$^*$  &  no nucleus      &    -    &  -  & -  &  -  &  - \\
N289$^*$  &   19.8(0.6)     &     0.14(0.03)   &  17.3(0.2)  &  0.14(0.06) &  16.7(0.2)  &  0.13(0.03)\\ 
N406  &  21.6(0.3)  &  0.04(0.01) &  21.0(0.3)  &  0.07(0.02) &  20.9(0.3)  &  0.10(0.08)\\ 	 
N488  &   no nucleus  &  -      &  -      &  -      & -  & - \\
N772  &  no nucleus      &   -     & -  & - &  -  &  - \\
N972$^*$  &  ?  & ? &  ? &  ? &  ?  &  ?\\ 	 
N986  &  16.0(0.4)  &  0.18(0.05) &  -      &    -    &  16.5(0.5)  &  0.11(0.01)\\
N1345 &  21.1(0.3)  &  0.05(0.01) &   -     &   -     &  20.4(0.1)  &  0.05(0.02)\\	 
N1398 &  no nucleus      &   -     &  -  & -  & -   &  -\\
N1483 &  21.8(0.2)  &  0.02(0.02)$\dagger\dagger$     &  -      &   -     &  21.6(0.5)  &  0.03(0.02)\\ 
N1688 &  19.6(0.3)  &  0.11(0.05) &    -    & -       &  19.2(0.8)  &  0.06(0.04)\\ 	 
      &  19.8(0.3)  &  0.12(0.05) &    -    & -       &  19.8(0.7)  &  0.03(0.03)\\
N1800 &   no nucleus     &  -      &  -    &  -      &  -  &  -     \\
N1892 & 21.2(0.7)       &    0.07(0.01)    &   -     &  -      &  18.5(0.3)  &  0.05(0.01)\\ 	 
N2082 &  21.0(0.3)  &  0.08(0.01) &  19.8(0.5)  &  0.08(0.01) &  19.4(0.7)  &  0.06(0.01)\\
N2196 &   no nucleus     &   -     &  -  &  -  & -  &  -\\
N2339 & ?      & ?       & ?     & ?     &   ?  & ?     \\ 
N2344 & no nucleus       &   -     &  -      &     -   & -  & - \\
N2460 &  no nucleus      &  -      &   -     &   -     &   - &  -\\  
N2566 &   -     &  -      &  14.6(0.5)  &  0.12(0.03) &  14.3(0.7)  &  0.11(0.01)\\ 	 
N2748 &    ?    &   ?     &    ?    &   ?     &  ?  &  ? \\
N2758 &  20.4(0.3)  &  0.11(0.05) &   -     &   -     &  20.2(0.6)  &  0.07(0.01)\\ 	 
N2964 &  18.4(0.5)  &  0.12(0.04)       &   -     &  -      &  16.2(0.4)  &  0.14(0.04)\\ 
N3067 &   ?     &   ?     &   ?     &   ?     &   ?     &   ?     \\
N3259&  19.5(0.1)  &  0.02(0.02)$\dagger\dagger$     &   -     &    -    &  18.4(0.7)  &  0.11(0.08) \\
N3277 &  17.3(0.3)  &  0.17(0.01) &  16.0(0.2)  &  0.14(0.02) &  15.7(0.2)  &  0.15(0.02) \\    
N3455 &  20.2(0.2)  &  0.07(0.01) &  19.2(0.3)  &  0.10(0.01) &  18.9(0.4)  &  0.04(0.01) \\   
N3898 & no nucleus       &  -      &     -   &    -    &  -  &  - \\    
N3900 &   no nucleus     &   -     &  -      &   -     &   - &  - \\    
N3928 &  17.1(0.3)  &  0.10(0.05) &   -     &   -     &  16.6(0.2)  &  0.10(0.01) \\    
N3949 &  -      &     -   &  18.0(0.3)  & 0.04(0.03) &  17.8(0.5)  &  0.05(0.03) \\    
N4102 &  ?      &   ?     &   ?     &   ?     &   ?     &   ?     \\    
N4219 &   ?     &   ?     &   ?     &    ?    &  ?  &  ? \\    
N4384 &  21.78(0.3)  &  0.08(0.03) &    -    &  -      &  21.5(0.8) & 0.06(0.04) \\
N4527 &  -      &   -     &   ?     &   ?     &   ?     &   ?     \\   
N4750 &  18.0(1.0)$^\dagger$  &  0.11(0.05)$^\dagger$   &  16.8(0.2)  &  0.15(0.03) &16.3(0.3)&0.14(0.03) \\
N4806 &  20.9(0.2)  &  0.07(0.01) &  19.9(0.4)  &  0.11(0.02) &  19.4(0.5)  &  0.05(0.01) \\    
N4980 &  20.3(0.2)  &  0.07(0.02) &  20.0(0.2)  &  0.07(0.01) &  19.7(0.3)  &  0.05(0.01) \\    
N5188 & 19.3(0.3)   & 0.08(0.01)  & 17.2(0.3)   & 0.24(0.12)  & 16.6(0.1)   & 0.20(0.09) \\     
N5377 &  16.3(0.2)  &  0.26(0.01) &  15.7(0.7)  &  0.19(0.02) &  15.8(0.2)  &  0.18(0.03) \\    
N5448 & ?  & ?  & -  & -  & ?  &  ? \\    
N5678$^*$ &  ?     &  ?      &  16.8(0.2)  &  0.21(0.03) &  16.4(0.3)  &  0.18(0.06) \\    
N5806 &  18.7(0.5)  &  0.15(0.02) &  16.7(0.3)  &  0.22(0.06) &  16.3(0.3)  &  0.21(0.06) \\    
N5985 &  ?      &     ?   &  ?  &  ? &  ?  &  ? \\    
N6000 &  17.6(0.4)  &  0.21(0.07)       &   -     &    -    &  16.6(0.4)  &  0.17(0.05) \\    
N6239 &   20.96(0.5)$^\dagger$ & 0.03(0.02) $^\dagger$  & 19.4(0.5)  & 0.05(0.02)     &  18.6(0.6)      &  0.03(0.03)   \\    
N6340 & no nucleus       &    -    &   -     &  -      &  - &  - \\    
N6384 &  19.7(0.6)  &  0.10(0.02) &  17.9(0.7)  &  0.12(0.04) &  17.6(0.8)  &  0.14(0.05) \\    
N6810 & saturated       &   -     &   ?  &    ?   &   ?    & ?     \\    
N6951 & -       & -        &  16.4(0.2)  &  0.17(0.04) &  16.0(0.3)  &  0.18(0.06) \\    
N7013 &    no nucleus    &   -     &  -      &   -     &  -   &  - \\    
N7162$^*$ &  21.0(0.1)     &   0.11(0.04)     &  19.9(0.4)  &  0.11(0.02) &  19.8(0.6)  &  0.11(0.03) \\    
N7177 &    -    &   -     &   ?     &    ?    &   ?     &   ?     \\    
N7188$^*$ &    19.9(0.4)     &  0.10(0.04)     &  19.3(0.3)  &  0.12(0.01) &  18.9(0.3)  &  0.06(0.02) \\    
N7217 &  no nucleus  &  - &  -  & - & -   & - \\    
N7259$^*$ &    20.6(0.6)   &     0.09(0.02)   &  19.3(0.3)  &  0.07(0.01) &  18.8(0.3)  &  0.06(0.03) \\    
N7280 &   no nucleus     &   -     &   -     &   -     &  - &  - \\    
N7421 &  19.1(0.5)  &  0.07(0.03) &  17.9(0.5)  &  0.09(0.01) &  17.6(0.8)  &  0.11(0.02) \\    
N7513 &  -      &    -    &  18.6(0.3)  &  0.12(0.02) &  18.3(0.4)  &  0.06(0.01) \\    
N7690 &  18.1(0.7)  &  0.09(0.01) &  16.8(0.3)  &  0.13(0.08) &  16.7(0.3)  &  0.17(0.05) \\    
\hline
\end{tabular}\end{center}}
\caption{Apparent magnitudes ($m_{band}$) and half-light radii
($R_{band}$) for the nuclei in the $V$, $J$ and $H$ bands. The optical
measurements are included for easy reference; they are extracted from
Carollo et al.\ (1997, 1998) for all galaxies but for the eight
identified by an asterisk (see caption Table 1, and text). Galaxies
observed with {\tt NICMOS} that do not host a photometrically-distinct
nucleus are also listed for completeness (and identified by `no
nucleus' in the second column). Question marks (?) identify systems in
which the presence of a nucleus is unclear. A dagger ($\dagger$)
in column 3 identifies the two nuclei for which  $V$ measurements of their 
sizes and
magnitudes were possible only after analysing the NIR images, which
revealed the location of the true center/nucleus. A double-dagger
($\dagger\dagger$) identifies nuclei that are unresolved on the PC of
{\tt WFPC2}. }
\label{tab2}
\end{table*}
\normalsize

\newpage

\begin{table*}
{\tiny\begin{center}\begin{tabular}{lcccll}
\hline\hline
\multicolumn{1}{l}{Name}  &
\multicolumn{1}{c}{Centered?} &
\multicolumn{1}{c}{Elongated?} &
\multicolumn{1}{c}{Color?}&
\multicolumn{1}{c}{Bar?} &
\multicolumn{1}{c}{Comments} \\
&&&&&\\
\hline \tiny
E205G7	&Y		&N	&B	&L		&red (dust) ring at $\sim 1''$ radius; red (dust) lane down to nucleus\\
E240G12	&?		&?	&N	&?		&round nucleus embedded in fainter elongated structure; \\
                                                            &&&&&several blue features (major one is shock-front-like, East of nucleus)\\
E404G3	&?		&N	&B	 &N?, L	        &inner elongated structure is warped\\
E498G5	&Y		&N	&?	&N?		&blue star forming spiral arms and red (dust) spiral-like  lanes\\
                                                            &&&&&over smooth round bulge\\
E499G37	&Y		&N	&B	&S		& Faint star forming spiral arm reaches smooth round `bulge' at $\sim 1''$\\
E548G29	&?		&N	&?	&L		&red (dust) patches and blue features down to nucleus\\
E572G22	&?		&N	&N	&L?		&NE  feature with blue and red knots\\
IC4390	&Y		&N	&N?	&S?, L		&red (dust) lane across nucleus; nuclear bar or disk\\
IC5271	&Y		&?	&B?     &S?, L	        &red bar/lane feature NE of nucleus\\
IC5273	&-		&-	&-	&S?, L	&multiple-knots, elongated  central blue feature  (not considered as  `nucleus');\\
                                                            &&&&&red (dust) patch NE of center\\
NGC289	&?		&?	&?      &S?, L?		&red (dust) lane across (NW of red) center \\
NGC406	&N		&N	&B?	&N		& red (dust) and blue star forming knots down to nucleus;\\
                                                            &&&&&triple blue structure $\approx10''$ NW of nucleus (background?)\\
NGC488	&-		&-	&-	&N		& red (dust?) spot at center; \\
&&&&&red (dust) spiral lanes down to center \\
NGC772  &-              &-      &-      &N            &red (dust) spiral lane down to blue center \\
NGC972	&?		&?	&?	&?		&central region obscured by dust \\
NGC986	&?		&?	&B	&L		&blue knots (on bar and spiral arms) and red (dust) lanes down to nucleus \\
NGC1345	&?		&?	&N	&L		&red (dust) patches down to nucleus; star forming knots \\
NGC1398	&-		&-	&-	&N		&center is blue; blue feature $\approx 4''$SE of center \\
NGC1483	&?		&?	&B	&?		&several cluster-like blue star forming knots in the nuclear region \\
NGC1688 &?		&?	&?	&L		&center not well defined even in $H$ \\
NGC1800 &?		&?	&?	&L?		&large number of cluster/star-type blue sources\\
NGC1892	&?		&?	&R	&N		&red (dust) lane across nucleus (reddened)\\
NGC2082	&?		&N	&R	&L		&red (dust) and blue (star forming) regions throughout central region \\
NGC2196	&-		&-	&-	&N		&spiral red (dust) feature SE of center (down to reddened center) \\
NGC2339	&?		&?	&?	&S?		&red center is clearly dust-reddened; blue source SW of center;\\
                                                            &&&&& blue-NE/red-SW oriented bar-like feature\\
NGC2344	&-		&-	&-	&N		&no major features in color map ($\sim0.2$ mag central reddening)\\
NGC2460	&-		&-	&-	&N		&blue center; blue features on spiral pattern down to center;\\
                                                            &&&&& red (dust) spiral features down to center\\
NGC2566	&Y		&N	&?	&N		&regular blue nuclear spiral arms; red (dust) lanes outside\\
NGC2748	&?		&?	&?	&N		&dust throughout central region\\
NGC2758	&N	&N	&N	&N	& faint red (dust) lanes through center\\
NGC2964	&Y		&Y	&N	&N		&spiral structure and red (dust) features down to center;  reddened nucleus?  \\
NGC3067	& ?		&?	&?	&N		& center has same color than surroundings; dust throughout central region \\
NGC3259	&Y		&N	&B	&N		&flocculent nuclear blue spiral arms and red (dust) lanes\\
NGC3277	&Y		&N	&R	&N		&red (dust) spiral features down to reddened nucleus\\
NGC3455	&?		&N	&N	&S?		&faint red (dust) patches in central region; nuclear spiral arms (and bar?)\\
NGC3898	&-		&-	&-	&S?, L? 	&red (dusty) center; red nuclear elongated feature; \\
                                                            &&&&& larger scale SE-NW red (dust?) disk/barlike feature\\
&&&&& connects with red spiral feature NE of center  \\
NGC3900	&-		&-	&-	&N		&center likely reddened by dust; red (dust) patches surround center \\
NGC3928	&Y		&N	&B	&N		&small scale regular spiral structure centered on nucleus; \\
                                                            &&&&& red/ (dust) feature surrounds nucleus\\
NGC3949	&Y		&N	&B	 &S?	& nuclear barlike feature (?);\\
&&&&& red (dust) lane $\approx 0.7''$ N of nucleus\\
NGC4102	&?		&?	&?	&S?, L		&center likely obscured; red (dust) rings/arms with blue knots around center\\
NGC4219	&?		&?	&?	&S?		&center likely obscured by spiral dust feature;\\
                                                            &&&&&nuclear star forming ring (or bar/arm)\\
NGC4384	&?		&?	&N	&S		&irregular red (dust) patterns down to center (nucleus obscured/reddened?)\\
NGC4527	&?		&?	&?	&L?		&dust (from disk) obscures center; likely no nucleus\\
NGC4750	&Y		&N	&?	&S, L		&red (dust) spiral features down to center (strongest NE of it)\\
NGC4806	&Y		&N	&R	&S?, L		&red (dust) lane across nucleus (reddened)\\
NGC4980	&N		&N	&R	&N		&granular appearance  \\
NGC5188 &?		&?	&?	&N		&dust in central region; location of center unclear;\\
                                                            &&&&&nucleus parameters in Table 2 given for \\
&&&&&blue feature N of central red (dust) patch\\
NGC5377	&Y		&Y	&B	&L		&regular NE/SW-oriented  red (dust) ring $\approx 5''$ from nucleus; \\
&&&&& inner ring-like red (dust) features \\
NGC5448	&Y		&N	&R?	&S		&red nuclear bar along nucleus orientation\\	
NGC5678	&Y		&?	&B? &S?		& nucleus embedded in mini-bar? nuclear red (dust) ring  around nucleus\\
NGC5806	&Y		&?	&?	&S		&clear bulge+bar structure in $H$ image (obscured by dust in $V$)\\
                                                            &&&&&red (dust) spiral structure across nucleus\\
NGC5985	&-		&-	&-	&L		&red (dust) lanes surround center\\
NGC6000	&?		&?	&B	&L		&nucleus surrounded by irregular red (dust) ring; \\
&&&&&red (dust) spiral features down to ring\\
NGC6239	&Y?		&?	&R	&L?		&irregular dust/star-formation\\
NGC6340 &-              &-      &-      &N             & red (dust) disk $\approx 0.6''$ diameter surrounds center;\\
&&&&& disklike red (dust) lane SW of center on slighly larger scales\\
NGC6384 &Y              &N      &R      &L?              &spiral/ringlike red (dust) feature down to dusty nucleus\\
NGC6810 &?		&Y	&B	&S?		&$V$ image is saturated; nucleus bluer than in $V-H$ color map;\\
&&&&& ubiquitous dust in central region\\
NGC6951	&Y		&N	&B? &N		& blue ring surrounds red center; red (dust) spiral lanes outside blue ring\\
NGC7013 &-              &-      &-      &N             &ubiquitous dust from disk  in central region (including center)\\
NGC7162	&Y		&?	&B?	&S?	& NE-SW red bar (or ring/disk) feature; red (dust) lane E of nucleus \\
NGC7177 &?              &?     &?          &S?, L           &red (dust) disk/spiral-like  lanes across center; blue large-scale bar \\
NGC7188	&?		&Y	&N	&N		& dust NW of nucleus\\
NGC7217 &-              &-     &-          &N             &blue disklike feature N of red center; blue `arm' (strongest patch $\approx 5''$ from center)\\
NGC7259	&?		&?	&R	&N		& nucleus embedded in (elongated?) red feature; nuclear  red (dust) lanes \\
NGC7280	&-		&-	&-	&S, L		&reddened center; red (dust) ringlike feature W of minor axis \\
NGC7421	&Y		&N	&B?	&L		&patchy red (dust) lanes down to nucleus\\
NGC7513	&Y		&N	&R	&N	& irregular circumnuclear dust\\
NGC7690	&Y		&?	&?	&S?		& red (dust) disk surrounds blue disk/bar feature (central `hole' in disk?) \\
\hline
\end{tabular}\end{center}}
\caption{\tiny Notes on individual galaxies. The column `Centered?' lists
whether the photometrically-distinct nucleus -- when present -- is
located (Y) or not (N) on the dynamical center of the host galaxy.
The column `Elongated?' lists whether the nucleus is round (N) or
elongated (Y) in projection.  The column `Color?' identifies nuclei
which are bluer (B) or redder (R) than their surroundings. An `N' in
this column indicates that the nucleus has a `normal' color (i.e.,
similar to the circumnuclear region). The information is extracted
from the $V-H$ color maps or, for the galaxies observed only with {\tt
NICMOS}, from the $J-H$ color maps.  The column `Bar?'  lists whether
a large- (L) or small- (S) scale bar is present in the system.
Question marks indicate uncertain detections due to a very complex
central structure. \normalsize}
\label{tab3}
\end{table*}
\normalsize

\newpage

\begin{table*}
{\tiny\begin{center}\begin{tabular}{lccccccc}
\hline\hline
\multicolumn{1}{l}{Name} &
\multicolumn{1}{c}{$D$} &
\multicolumn{1}{c}{Type} &
\multicolumn{1}{c}{$m_B$} &
\multicolumn{1}{c}{$A_B$} &
\multicolumn{1}{c}{$m_H$} & 
\multicolumn{1}{c}{$R_H$} \\
&(Mpc)&&(mag)&(mag)&(mag)&($''$)\\
\hline \tiny
NGC1365 & 26 &.SBS3..& 10.32 & 0.00& 14.1(0.2)&0.04(0.01)\\
NGC2903 & 9 &.SXT4..& 9.61  & 0.07& 17.8(0.4) &0.14(0.04)\\
NGC3031 & ?  &.SAS2..& 7.75  & 0.14& 14.9(0.3)&0.05(0.02)\\
NGC4536 & 29 &.SXT4..& 11.16 & 0.00& ?&?\\
NGC4565 & 18 &.SAS3..& 10.39 & 0.03&? &?\\
NGC4569 & ? &.SXT2..& 10.17 & 0.09&15.2(0.5) &0.09(0.02)\\
NGC6217 & 21 &RSBT4..& 11.65 & 0.14& 17.2(0.3)&0.12(0.04)\\
NGC7479 & 37 &.SBS5..& 11.44 & 0.16&- &-\\
NGC7742 & 26 &.SAR3..& 12.25 & 0.10&- &-\\
\hline
\end{tabular}\end{center}}
\caption{The nine galaxies for which only F160W images are available.
The columns give the distance in Mpc (derived from the RC3 recession
velocity; for NGC3031 and NGC4569 the RC3 lists -34 and -331 km/s,
respectively), the morphological type and the apparent total B
magnitude from the RC3,
and the $B$ extinction from Burstein \& Heiles (1984). The last two
columns list the $H$ magnitudes and the half-light radii for their
nuclei (when present).}
\label{tab4}
\end{table*}
\normalsize